\documentstyle[aps,floats,epsf]{revtex}
\begin{document}                
\setlength{\epsfxsize}{4.0in}
\title{RADIATION FROM VIOLENTLY ACCELERATED BODIES\footnote{Published in Phys. Rev. D\, {\bf64}, 105004 (2001)}}

\author{ULRICH H. GERLACH}
\address{Department of Mathematics, Ohio State University, Columbus, OH
43210, USA}
\maketitle

\begin{abstract}
A determination is made of the radiation emitted by a linearly
uniformly accelerated uncharged dipole transmitter. It is found that,
first of all, the radiation rate is given by the familiar Larmor
formula, but it is augmented by an amount which becomes dominant for
sufficiently high acceleration. For an accelerated dipole oscillator,
the criterion is that the center of mass motion become relativistic
within one oscillation period. The augmented formula and the
measurements which it summarizes presuppose an \emph{expanding}
inertial observation frame.  A \emph{static} inertial reference frame
will not do.  Secondly, it is found that the radiation measured in the
expanding inertial frame is received with 100 percent fidelity. There
is no blueshift or redshift due to the accelerative motion of the
transmitter.  Finally, it is found that a pair of coherently radiating
oscillators accelerating (into opposite directions) in their
respective causally disjoint Rindler-coordinatized sectors produces an
interference pattern in the expanding inertial frame. Like the pattern
of a Young double slit interferometer, this Rindler interferometer
pattern has a fringe spacing which is inversely proportional to the
proper separation and the proper frequency of the accelerated sources.
The interferometer, as well as the augmented Larmor formula, provide a
unifying perspective. It joins adjacent Rindler-coordinatized
neighborhoods into a single spacetime arena for scattering and
radiation from accelerated bodies.

\end{abstract}

\section{INTRODUCTION}
The emission or the scattering of light from localized sources is the
most effective way for information to be transferred to the human eye,
the window to our mind. We can increase the size of this window, with
specialized detectors, or systems of detectors. They intercept the
information, record and re-encode it, before passing it on to be
assimilated and digested by our consciousness.

The spacetime framework for most physical measurements, in particular
those involving radiation and scattering processes, consists of
inertial frames, or frames which become nearly inertial by virtue of
the limited magnitude of their spatial and temporal extent. Indeed,
the asymptotic ``in'' and ``out'' regions of the scattering matrix as
well as the asymptotic ``far-field'' regions of a radiator reflect the
inertial nature of the spacetime framework for these processes.

Should one extend these processes to accelerated frames? If so, how?
Let us delay answering the first question and note that Einstein, in
his path breaking 1907 paper \cite{Einstein1907}, gave us the answer
to the second: View an accelerated frame as a sequence of
instantaneous locally inertial frames. Thus a scattering (or any other
physical) process observed relative to a lattice of accelerated clocks
and equally spaced detectors can be understood in terms of the lattice
of inertial clocks and equally spaced inertial detectors
\cite{detectors} of one or several of these instantaneous locally
inertial frames. Accelerated frames seem to be conceptually
superfluous!  Acceleration can always be transformed away by replacing
it with an appropriate set of inertial frames. To make observations
relative to an accelerated frame comprehensible, formulating them in
terms of a sequence of instantaneous inertial frames seems (at first
sight) to be sufficient.

The introduction of these inertial frames into physics was one of the
two historical breakthroughs \cite{Einstein1907b} for Einstein,
because mathematically they are the tangent spaces, the building
blocks from which he built general relativity.

However, characterizing an accelerated frame as a one-parameter family
of instantaneous Lorentz frames was only an approximation, as Einstein
himself points out explicitly\cite{Einstein1907} in his 1907
article. The approximation consists of the fact that the Lorentz
frames \emph{never have relativistic velocities} with respect to one
another. Thus Einstein approximated a hyperbolic world line in $I$ of
Figure 1 by replacing it with a finite segment having the approximate
shape of a parabola.  If Einstein had not made this assumption, then
he would have found immediately that associated with every uniformly
linearly accelerated frame there is a \emph{twin} moving into the
opposite direction, and causally disjoint from the first.  Nowadays
these twins are called Rindler sectors $I$ and $II$ as in Figure 1.
\begin{figure}[h!]
\epsfclipon
  \setlength{\epsfysize}{3.5in}
  \setlength{\epsfxsize}{6.0in}
\centerline{\epsffile[0 400 500 680]{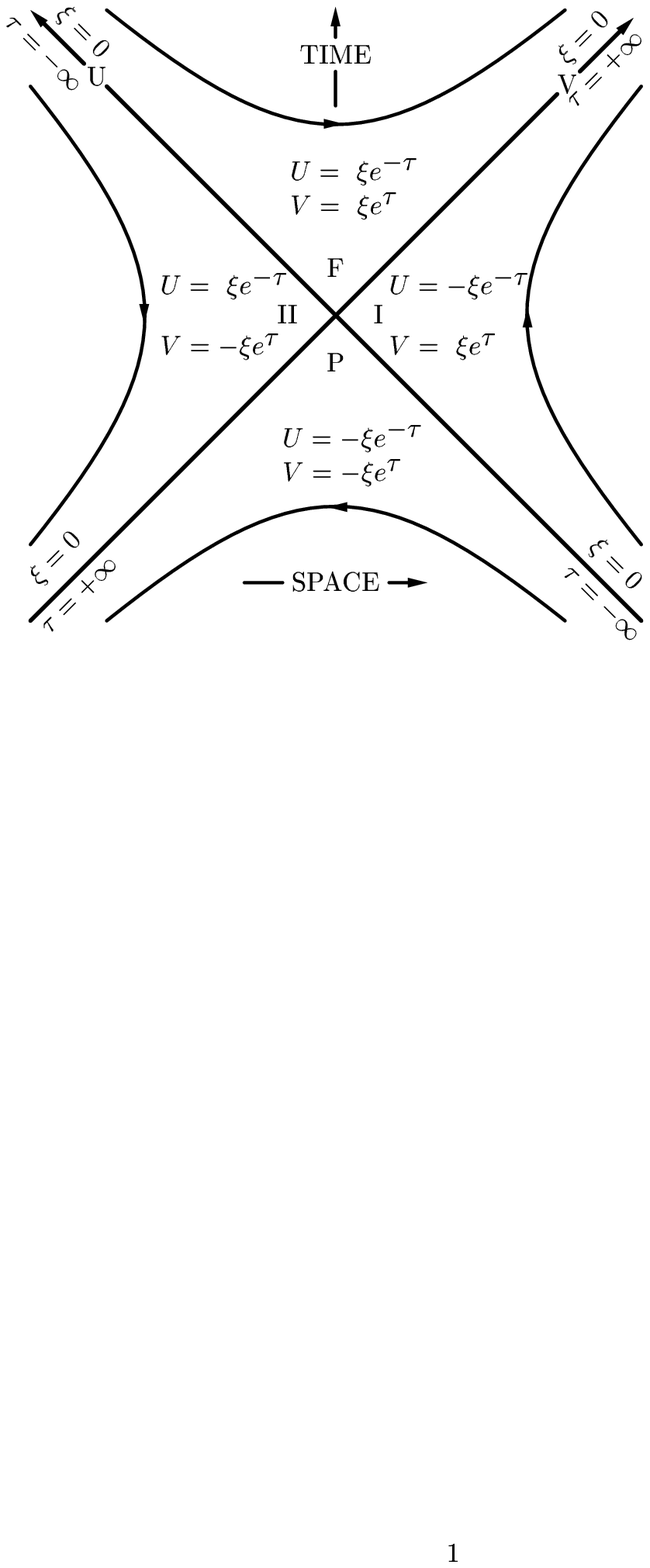}}
       \epsfverbosetrue
\caption{ Acceleration-induced partitioning of spacetime into the four
Rindler coordinatized sectors. They are centered around the reference event
$(t_0,z_0)$ so that $U=(t-t_0)-(z-z_0)$ and $V=(t-t_0)+(z-z_0)$ are 
the retarded and advanced time coordinates for this particular quartet
of Rindler sectors.}
\label{fig:Rindler spacetime}
\end{figure}
Thus, \emph{linearly uniformly accelerated frames always come in pairs},
which (a) are causally disjoint and (b) have lightlike boundaries, their
past and future horizons.

These two Rindler coordinatized sectors together with their past $P$
and future $F$ form a \emph{double-slit
interferometer}\cite{Gerlach1999} relative to a spatially homogeneous
but expanding coordinate frame. The two Rindler sectors $I$ and $II$
comprise the double slit portal through which wave fields propagate
from $P$ to $F$. During this process the wave field interacts with
sources, which due to their acceleration, are confined to, say,
Rindler sectors $I$ and/or $II$. The interference between the waves
coming from these two sectors is observed in $F$. There the field
amplitude is sampled in space and in time.

Consider the field which is due to accelerated sources in $I$ or in
$II$. A single inertial radio receiver which samples the field
temporally is confronted with a metaphysically impossible task: Track
and decode a signal with a Doppler chirp (time dependent Doppler
shift) whose phase is logarithmic in time. The longer and more violent the
acceleration of the source, the more pronounced the initial blueshift
and/or the final redshift at the receiver end. Tracking the amplitude
and the phase of such a chirped signal becomes a debilitating task for
any receiver.

Suppose, however, the field gets intercepted by a set of mutually
receding radio receivers. If they, in concert, sample the field
spatially at a single instant of ``synchronous'' time, then there is
no Doppler chirp whatsoever. An accelerated source which emits a sharp
spectral line will produce an equally sharp spectral line in the
spatial Fourier domain of the sampled space domain (in Figure
\ref{fig:Rindler spacetime}: $UV=\xi^2=const.$) of the expanding set
of radio receivers. In brief, \emph{a signal emitted by an accelerated
point source is intercepted by a set of mutually receding phased radio
receivers with 100 \% fidelity}. We shall refer to this result as the
\emph{fidelity property} of Rindler's spacetime geometry.

The physical reason for this result is given in Section
\ref{sec-fidelity:physical}, the mathematical formulation in Section
\ref{sec-fidelity:mathematical}.

The application of the fidelity property to the power emitted from an
accelerated dipole oscillator is given in Section \ref{sec-radiated
power}. This application consists of Larmor's formula\cite{Landau1962}
augmented due to the fact that the oscillator is in a state of
uniform acceleration.

The fidelity property applies to the radiation from a source
accelerated in Rindler $I$ as well as to a source accelerated in
Rindler $II$. If the two sources have the same frequency and are
coherent, then the phased array of radio receivers measures an
interference pattern which is mathematically indistinguishable from
that due to a standard double slit. This result is spelled out in
Section \ref{sec-double slit}.

It is worth while to reiterate that the fidelity property and its two
applications are statements about the Rindler coordinate neighborhoods
considered jointly, with the event horizons, $\xi=0$, integral
building blocks of these concepts. 
Some workers in the field \cite{MTWch1}, who view spacetime only in
terms of ``coordinate patches'' or ``coordinate charts'' (i.e. comply
with Einstein's approximation mentioned above), tend to compare the
locus of events $\xi=0$ in Fig.~\ref{fig:Rindler spacetime} to the
coordinate singularity at the North Pole of a sphere or the origin of
the Euclidean plane. Such a comparison leads to a pejorative
assessment of Rindler's coordinatization as
``imperfect'',``singular'', or ``poor'' at $\xi=0 $\cite{MTWch1}.
This is unfortunate. As a result, this comparison diverts attention
from the fact that (1) waves from $I$ and $II$ interfere in $F$ and
that (2) as a consequence, the resulting interference patterns serve
as a natural way of probing and measuring scattering and/or radiative
sources as well as gravitational disturbances in regions $I$ and $II$.

The Rindler double-slit opens additional vistas into the role of
accelerated frames $I$ and $II$. They accommodate causally disjoint
but correlated radiation and scattering centers whose mutually
interfering radiation is observed and measured in $F$. These
measurements are mathematically equivalent to having two accelerated
observers in Rindler $I$ and $II$ respectively. From these
measurements one can reconstruct in all detail the location and
temporal evolution of all accelerated radiation sources. The aggregate
of these sources comprises what in Euclidean optics is called an
object, one in Rindler $I$ the other in Rindler $II$. What is observed
in $F$ is the interference of two coherent diffraction patterns of
these two objects.  These measurements are qualitatively different
from those that can be performed in any \emph{static} inertial frame. They
yield the kind of information which can be gathered only in
accelerated frames with event horizons. One of the virtues of the
Rindler double-slit interferometer is that it quite naturally avoids
an obvious metaphysical impossibility\cite{MTW1973a}, namely, have
accelerating observers in Rindler sectors $I$ and $II$ which (a) have
the physical robustness to withstand the high (by
biological-technological standards) acceleration and/or (b) the
longevity and the propulsion resources to co-accelerate for ever and
never cross the future event horizon.

From the perspective of implementing measurements, the Rindler
double-slit has advantages akin to those of a Mach-Zehnder
interferometer \cite{Born and Wolf}: it permits an interferometric
examination of regions of spacetime whose expanse is spacious enough
to accommodate disturbances macroscopic in extent, and it permits one
to achieve this feat without putting the measuring apparatus into
harm's way.  However, in order to use the Rindler interferometer as a
diagnostic tool one must first have the necessary conceptual
infrastructure.  This article provides four of its ingredients:
\begin{itemize}
\item
Expanding free float frame
\item
Fidelity property of the Rindler spacetime geometry
\item
Augmented Larmor formula
\item
Double slit interference due to a pair of accelerated sources
\end{itemize}
\emph{Nomenclature:} This articlae uses repeatedly the words ``Rindler
sector'', ``Rindler spacetime'', etc. This is verbal shorthand for
``Rindler coordinatized sector'', ``Rindler coordinatized spacetime''
etc. The implicit qualifier ``coordinatized'' is essential because,
without it, ``Rindler sector/spacetime'' would become a mere floating
abstraction, i.e. an idea severed from its observational and/or
physical basis.

\section{EXPANDING INERTIAL OBSERVATION FRAME}

Fundamentally all of physics, including the physics of spacetime, is
based on measurements. The class of measurements we shall focus on are
those made by ``recording clocks'' \cite{Taylor and Wheeler 1992} in a
state of ``free float''. The meaning of a ``recording clock'' is that
each one of them consists of
\begin{itemize}
\item[(i)] a clock oscillator which controls the clock and has a
standard frequency,
\item[(ii)]
a transmitter whose emission frequency is controlled by this oscillator,
\item[(iii)]
a receiver capable of measuring the emitted radiation from the other 
recording clocks, even if there is a Doppler shift, and
\item[(iv)]
memory chips which can hold data acquired by the receiver.
\end{itemize}
Thus each clock is constructed like one of the GPS (Global Positioning
System) units orbiting the earth.  Assuming no gravitation, one says
that the aggregate of recording clocks is in a state of ``free float''
(inertial motion) if the relative Doppler shift between each pair of
such clocks is fixed and constant in time. There is \emph{no Doppler
chirp}. In other words, each recording clock measures (and stores in
its memory) spectral lines which are sharp. The sharper the measured
spectral lines the more closely the recording clocks are in a state of
free float.

From this swarm of freely floating recording clocks one now forms, by a
process of measurement omission, an equivalence class called an
``expanding free float frame''. 

\subsection{Construction}

The formation of this concept is achieved by using one of the clocks,
say $R$, as a reference clock which measures and collects two kinds of
data about all the other clocks, say $A,B,C,\cdots$ : Doppler shifts and 
instantaneous distance of $A,B,C,\cdots$ from $R$.

Doppler shift measurements are frequency measurements. The emission
frequencies of all clocks $A,B,C,\cdots$ are the same, say $\omega_0$.
Consequently, the frequencies $\omega_A,\omega_B,\omega_C,\cdots$
received and measured by $R$ yield the corresponding Doppler shift factors
\[
k_A=\frac{\omega_A}{\omega_0}\equiv \exp{(-\tau_A)},
~k_B=\frac{\omega_B}{\omega_0}\equiv \exp{(-\tau_B)},~
k_C=\frac{\omega_C}{\omega_0}\equiv \exp{(-\tau_C)},~ \cdots
\]
for the respective clocks. From these Doppler shift factors one obtains the relative velocities 
\[
v_A=\frac{\displaystyle\frac{1}{k_A}-k_A}{\displaystyle\frac{1}{k_A}+k_A}\equiv \tanh \tau_A,~
v_B=\frac{\displaystyle\frac{1}{k_B}-k_B}{\displaystyle\frac{1}{k_B}+k_B}\equiv \tanh \tau_B,~
etc.
\]
between $R$ and the respective clocks.

The second kind of measurement is the instantaneous separation.
Suppose at some instant of time, say $t_R$, $R$ measures the distances
$d_A,d_B,d_C, \cdots$ between $R$ and the clocks $A,B,C,\cdots$.
One way of doing this is to have $R$ operate his transmitter
and receiver as a radar device.

We now say that $A$ is equivalent to $B$, or more briefly $A\sim B$, if 
\begin{equation}
\frac{d_A}{v_A}= \frac{d_B}{v_B}\equiv \xi (t_R)~.
\label{eq:ratios}
\end{equation}
It follows that 
\begin{itemize}
\item[1.]
$A\sim B$ implies $B\sim A$ and 
\item[2.]  $A\sim B$ and $B\sim C$ imply $A\sim C$.  
\end{itemize}
By retaining the distinguishing property, the equality of the ratios,
Eq.(\ref{eq:ratios}), while omitting reference to the particular
measurements $d_A,d_B,d_C, \cdots$ and $k_A,k_B,k_C,\cdots$ one forms
an equivalence class, the concept ``expanding inertial frame''. It
should be noted that the reference clock $R$ is always a member of
this equivalence class. This is because, with $d_R=0$ and $v_R=0$, the
equivalence condition
\[
\frac{d_R}{v_R}= \frac{d_B}{v_B}
\]
is satisfied trivially.  

The properties of this equivalence class do not depend on the time at
which $R$ makes the distance measurements and hence not on the value
of the ratio $\xi(t_R)$. Indeed, if instead of $t_R$ that time had
been , say $t'_R$, then the corresponding distance measurements would
be $d'_A,d'_B,d'_C, \cdots$, then one would still have a set of equal
ratios
\[
\frac{d'_A}{v_A}= \frac{d'_B}{v_B}\equiv \xi (t'_R)~,
\]
which would yield the same equivalence class of free float clocks.

The purpose of an inertial reference frame is for a physicist/observer
to use its recording clocks to measure time and space
displacements. These measurements consist of establishing quantitative
relationships (typically via counting) to a standard which serves as a
unit. For a time measurement the unit is the standard interval between
any two successive ticks of a clock. For a space measurement the unit
is the (logarithm of the) standard Doppler shift factor between any
pair of nearest neighbor clocks. Thus the array of clocks forms a
lattice which is periodic but is expanding uniformly: the recession
velocity between any neighboring pair of clocks is one and the
same. This periodicity is an obvious but tacit stipulation in what is
meant by ``expanding inertial frame''. Because of this property any
one of the recording clocks $A,B,C,\cdots$ can play the role of the
reference clock $R$, which is to say that the equivalence relation,
Eq.(\ref{eq:ratios}), is independent of the choice of $R$.

The two kinds of measurements which gave rise to the equivalence
relation between recording clocks also serve to synchronize their
operations. Every clock synchronizes itself to its nearest neighbor by
setting its own clock reading to the ratio of (i) the nearest neighbor
distance and (ii) the Doppler shift determined velocity. Thus the
common ratio, Eq.(\ref{eq:ratios}), is the synchronous time common to
all recording clocks.  This common time has an obvious interpretation:
The straight-line extensions into the past of all clock histories
intersect simultaneously in a common point. This is a singular event,
which corresponds to $\xi=0$. The common synchronous time of these
clocks is the elapsed proper time since then.  However, it is obvious
that this singular event is irrelevant for the definition of the
expanding inertial frame. What is relevant instead is the ability of
the recording clocks to measure Doppler shifts and distances, which
presupposes that $\xi \neq 0$. In fact, these clocks might not even
have existed until they performed their measurements.

Having constructed the spacetime measuring apparatus, we indicate in
general terms how to make spacetime measurements of particles and
fields.

\subsection{Measurements of Particles and Fields}

The mechanical measurements by a physicist/observer of the
spacetime properties of a classical particle consists of (a) identifying which
clock detects the existence of the particle in what interval of
synchronous time and (b) determining the particle's velocity by
measuring its Doppler shift. The first is a counting process in space
and time, the second is a counting process in temporal frequency
space.

The wave mechanical measurements of the spacetime properties of a
classical electromagnetic (e.m.) field consist of using the expanding
set of recording clocks to form a \emph{phased array} of mutually
receding receivers.  This array samples the e.m. field at the
locations of the recording clocks at regular intervals of synchronous
time. The phased array mode involves all clocks at once and thus
provides a record of the magnitude and spatial phase of the
e.m. field. By repeating this procedure at temporal intervals
controlled by the synchronized ticking of the clocks, one obtains a
sampled historical record of the magnitude of the field and its
temporal phase.

\section{Transmission Fidelity}
\label{sec-fidelity:physical}

The transfer of information from a transmitter to a receiver, or a
system of receivers, depends on being able to establish a one to one
correspondence between (i) the phase and amplitude of the e.m. source
and (ii) the e.m. signals detected by the observer who mans the
receiver(s) in his frame of reference.  For a localized source with a
straight worldline that frame is \emph{static} and inertial. For a
source with a hyperbolic worldline as in Figure \ref{fig:Rindler
spacetime}, it is \emph{expanding} and inertial. The e.m. signals are
detected by having the recording clocks sample and measure the
e.m. field at any fixed synchronous time $\xi>0$. Except for a
$\xi$-dependent amplitude and domain shift, these measured field
values (along the spatial domain $-\infty<\tau<\infty$) are precisely
the values of the current source (along the temporal domain
$-\infty<\tau'<\infty$) of the accelerated
transmitter\cite{notation}. Geometrically one says that the
transmitter signal-function, whose domain is a timelike hyperbola in
Rindler sector $I$, coincides in essence with the receiver signal
function whose domain is a spacelike hyperbola in Rindler sector
$F$. Thus, if the signal-function is monochromatic at the transmitter
end, then so is the signal function on the spatial domain at the
receiver end. \emph{There is no chirp} (changing wave length) \emph{in
the spatial wave pattern in the expanding inertial frame.}

One arrives at that conclusion by verifying it for wave packets,
i.e. for narrow but finite pulses of nearly monochromatic radiation,
which make up the e.m. signal. Thus consider a uniformly and linearly
accelerated transmitter. The history of its center of mass is
represented by a timelike hyperbola in, say, Rindler sector $I$ in
Figure \ref{fig:Rindler spacetime}. 

Let us have this single transmitter emit \emph{two successive} pulses
which have the same mean frequency and require that they be received
at the same synchronous time $\xi$ by \emph{two adjacent} recording
clocks in $F$. One has therefore two well-defined emission-reception
processes,
\begin{equation}
(\tau'_A,\xi'):A(\tau'_A,\xi')
			\begin{array}{c}
				\textrm{world history of pulse }A\\
				\textrm{-------------------------}\!\!\!\longrightarrow\\
				\textrm{  } 
			\end{array} 
(\xi,\tau_A):A(\xi,\tau_A)
\label{eq:process A}
\end{equation}
and
\begin{equation}
(\tau'_B,\xi'):B(\tau'_B,\xi')
			\begin{array}{c}
				\textrm{world history of pulse }B\\
				\textrm{-------------------------}\!\!\!\longrightarrow\\
				\textrm{  }
			\end{array} 
(\xi,\tau_B):B(\xi,\tau_B)~.
\label{eq:process B}
\end{equation}
Each starts with a pulse emitted by the accelerated transmitter in
Rindler sector $I$ and ends with the pulse's reception by an inertial
recording clock in Rindler sector $F$. Both processes end with the
simultaneous reception of these pulses in the expanding inertial
reference frame. Among its recording clocks there are precisely two,
labelled by $\tau_A$ and $\tau_B$, which receive pulses $A$ and $B$.

The first process starts with pulse $A$ at event $(\tau'_A,\xi')$ on the
timelike hyperbolic world line $\xi'=const'.$ in Rindler sector $I$.
That pulse is launched from the instantaneous Lorentz frame
$A(\tau'_A,\xi')$ centered around this event. Having traced out its
world history across the future event horizon of $I$, this pulse ends
the first process at event $(\xi,\tau_A)$ on the spacelike hyperbola
of synchronous time $\xi=const.$ There, in the local Lorentz frame
$A(\xi,\tau_A)$ of the inertial recording clock with label
$\tau=\tau_A$, the (mean) wavelength of the pulse is measured and
recorded.

The second process starts with pulse $B$ emitted at event
$(\tau'_B,\xi')$ on the \emph{same} timelike hyperbola $\xi'=const'.$
but at \emph{different} Rindler time $\tau'=\tau'_B$. This pulse also
traces out a world history across the future event horizon. But the
end of this pulse is at $(\xi,\tau_B)$ on the \emph{same} spacelike
hyperbola $\xi=const.$ There the (mean) wavelength of the pulse gets
measured relative the local Lorentz frame $B(\xi,\tau_B)$ of the
inertial recording clock with \emph{different} label $\tau=\tau_B$.
Even though pulses $A$ and $B$ are emitted sequentially by one and the
same transmitter, they are received simultaneously by two different
recording clocks. This is made possible by the fact that the clock
labelled by $\tau_B$ is moving towards the approaching pulse $B$. The
blueshift resulting from this motion precisely compensates the
redshift which pulse $B$ has relative to $A$ \emph{if} the recording
clock did not have this motion. Thus recording clocks $\tau_A$ and
$\tau_B$ receive pulses $A$ and $B$ having precisely the same
respective frequencies.  This agreement is guaranteed by the principle
of relativity. Indeed, Eq.(\ref{eq:process A}) is a Lorentz transform
of (\ref{eq:process B}).  Each consists of two events, two sets of
frame vectors, and a straight pulse history. The Lorentz
transformation maps these five entities associated with pulse $A$ into
those associated with pulse $B$:
\begin{equation}
\begin{array}{ccccc}
(\tau'_A,\xi'):&A(\tau'_A,\xi')&
			\begin{array}{c}
				\textrm{world history of pulse }A\\
				\textrm{-------------------------}\!\!\!\longrightarrow\\
				\textrm{  } 
			\end{array} &
(\xi,\tau_A):&A(\xi,\tau_A)\\
\downarrow&\downarrow&\downarrow&\downarrow&\downarrow\\
(\tau'_B,\xi'):&B(\tau'_B,\xi')&
			\begin{array}{c}
				\textrm{world history of pulse }B\\
				\textrm{-------------------------}\!\!\!\longrightarrow\\
				\textrm{  }
			\end{array} &
(\xi,\tau_B):&B(\xi,\tau_B)
\end{array}
\end{equation}

Thus the relative velocity, and hence the Doppler shift between frames
$A(\xi',\tau'_A)$ and $A(\xi,\tau_A)$, is the same as that between
$B(\tau'_B,\xi')$ and $B(\xi,\tau_B)$. This means that the wavelengths
of the two received pulses at clock $A(\tau_A)$ and clock $B(\tau_B)$
are the same.  There is \emph{no Doppler chirp} in the composite
spatial profile of the received e.m. field at fixed synchronous time
$\xi$. If the emitted signal is monochromatic relative to the
accelerated transmitter in $I$, then so is the spatial amplitude
profile of the received signal relative to the expanding inertial
frame in $F$.  The transmission of a sequence of pulses is achieved
with 100\% fidelity.

This conclusion applies to all wavepackets. It also applies to any
signal.  This is because it is a linear superposition of such packets.
A precise mathematical formulation of the emission of signals and
their fidelity in transit from an accelerated source to an expanding 
inertial frame is developed in Section \ref{sec-fidelity:mathematical}.

Some authors thought that there is some sort of a disconnect between
mathematics and physics, in particular between computations and what
the computations refer to. For example, they claimed that ``$\cdots$
the coordinates that we use [for computation] are arbitrary and have
no physical meaning''\cite{Wigner1980} or ``It is the very gist of
relativity that anybody may use any frame [in his
computations].''\cite{Schroedinger1956} Without delving into the
epistemological fallacies underlying these claims, one should be aware
of their unfortunate consequences. They tend to discourage attempts to
understand natural processes whose very existence and identity one
learns through measurements and computations based on nonarbitrary
coordinate frames.  The identification of radiation from violently
accelerated bodies is a case in point. For these, two complementary
frames are necessary: an accelerated frame to accommodate the source
(Rindler sector $I$ and/or $II$) and the corresponding expanding
inertial frame (Rindler sector $F$) to observe the information carried
by the radiation coming from this source. These frames are physically
and geometrically distinct from static inertial frames. They also
provide the logical connecting link between the concepts and the
perceptual manifestations (measurements) of these radiation
processes. Without these frames the concepts would not be concepts but
mere floating abstractions.

\section{MAXWELL FIELDS: TRANSVERSE ELECTRIC AND TRANSVERSE MAGNETIC}
\label{sec-Maxwell fields}
The connecting link between the observed electromagnetic (e.m.) field
and its source is the Maxwell field equations. The linear acceleration
of the source, as well as the relative motion of the expanding set of
recording clocks, determine the axis of a cylindrical geometry. This
geometry results in the solutions to the e.m. field being decomposed
into two distinct independently evolving fields the familiar
T.M. polarized field and the T.E. field.  Each is based on a single
scalar field, which is a scalar under those Lorentz transformations
which preserve cylindrical symmetry. In particular, the
T.M. (resp. T.E.) field has vanishing magnetic (resp. electric) but
nonzero electric (resp. magnetic) field parallel to the cylinder
axis. Finally, there are also the T.E.M. fields. They are both
T.E. and T.M. at the same time, and they propagate strictly parallel
to the cylinder axis.

There is an analogous T.M.-T.E.-T.E.M. decomposition of the
source. For example, the difference between the T.M. and the
T.E. fields is that the source of the T.M. fields is the density of
electric multipoles, while the source for the T.E. fields is the
density of magnetic multipoles.

\subsection{The Method of the 2+2 Split}

The first task is to solve the inhomogeneous Maxwell field
equations\cite{Candelas plus Higuchi,Alexander and Gerlach} and use
its solution to determine the radiation properties to be measured.
The best way to set up and solve these equations is to take advantage
of the fact the cylindrical symmetry dictates a 2+2 decomposition of
spacetime into a pair orthogonal 2-dimensional planes, one Euclidean,
the other Lorentzian. The experienced reader will see that such a
decomposition minimizes (compared to text book treatments) the amount
of mathematical analysis, while simultaneously retaining all physical
aspects of the radiation problem. Furthermore, these physical aspects
lend themselves to nearly effortless identification because of the
flexible curvilinear coordinate systems which the two orthogonal
planes accommodate.

However, in order to appreciate the underlying line of reasoning as
rapidly as possible, we first illustrate the 2+2 decomposition
procedure on Minkowski spacetime coordinatized with the familiar
rectilinear coordinates. We do this before proceeding to use it to
solve Maxwell's equations relative to the various Rindler coordinates
plus polar coordinates as called for by the cylindrical
coordinate geometry of linearly accelerating bodies.

In the presence of cylindrical symmetry the Maxwell field equations
decouple into two sets, each of which gives rise to its own inhomogeneous
scalar wave equation
\begin{equation}
\left[
	\left(
-{\partial^2  \over \partial t^2}
+{\partial^2  \over \partial z^2}
	\right) +
	\left(
{\partial^2  \over \partial x^2}+
{\partial^2  \over \partial y^2}
	\right)
\right]
\psi(t,z,x,y) =-4\pi\, S(t,z,x,y)~~.
\label{eq:the full equation}
\end{equation}
The charge-current four-vector 
\begin{equation}
S_\mu dx^\mu=S_t dt +S_z dz +S_x dx+S_y dy
\label{eq:rectilinear charge current four-vector}
\end{equation}
for each set determines and is
determined by appropriate derivatives of the scalar source
$S(t,z,x,y)$. Similarly, 
the vector potential
\begin{equation}
A_\mu dx^\mu =A_t dt+A_z dz+ A_x dx +A_y dy
\label{eq:the vector potential}
\end{equation}
as well as the electromagnetic field 
\begin{eqnarray*}
\frac{1}{2}F_{\mu\nu}dx^\mu \wedge dx^\nu 
&\equiv&
	\frac{1}{2}(\partial_\mu A_\nu -\partial_\nu A_\mu)dx^\mu \wedge dx^\nu\\
&\equiv&
\hat E_{long.}dz \wedge dt +\hat E_y dy\wedge dt 
							+\hat E_x dx\wedge 
						dt\\
&~&~+\hat B_xdy\wedge dz +\hat B_y dz\wedge dx +\hat B_{long.}dx\wedge dy 
\end{eqnarray*}
determine
and are determined by the scalar wave function $\psi(t,z,x,y)$
with the result that 
the Maxwell field equations
\begin{equation}
{F_\mu^{~\nu}}_{;\nu}=4\pi S_\mu
\label{eq:Maxwell's equations}
\end{equation}
are satisfied whenever the wave Eq.(\ref{eq:the full equation})
is satisfied.

\subsubsection{The T.E. Field}
\label{The T.E. Field}

For the T.E. degrees of freedom the components of the charge-current 
four-vector are
\begin{equation}
( S_{t}, S_{z}, S_{x}, S_{y})=\left(0,0,
\frac{\partial S}{\partial y},
-\frac{\partial S}{\partial x}\right)~.
\label{eq:rectilinear T.E. source}
\end{equation}
The components of the T.E. vector potential are
\begin{equation}
( A_{t}, A_{z}, A_{x}, A_{y})=\left(0,0,
\frac{\partial \psi}{\partial y},
-\frac{\partial \psi}{\partial x}\right)~,
\label{eq:rectilinear T.E. vector potential}
\end{equation}
and those of the e.m. field are 
\begin{center}
\begin{tabular}[t]{|l|c|}	
\hline
$ E_{long.}:$
& $  F_{zt}=0 $\\ 
								[0mm]\hline
$ E_x:$&$ \displaystyle  F_{xt}=
 -\frac{\partial }{\partial y}\frac{\partial \psi}{\partial t } $\\
								[3mm]	\hline
$ E_y:$  &  $\displaystyle F_{yt}=
\frac{\partial }{\partial x}\frac{\partial \psi}{\partial t} $
								\\[3mm]\hline
$ B_{long.}:$ & $\displaystyle F_{xy}=
-\left( 
\frac{\partial^2}{\partial x^2}
 +\frac{\partial^2}{\partial y^2} \right) \psi $\\	[2mm]\hline
$ B_x:$ & $\displaystyle F_{yz}=
\frac{\partial }{\partial x}
 \frac{\partial \psi }{\partial z}	$\\
								[3mm]\hline
$ B_y:$ & $\displaystyle F_{zx}=
\frac{\partial }{\partial y}
		\frac{\partial \psi}{\partial z}$\\
								[3mm]\hline
\end{tabular}
\end{center}
These components are guaranteed to satisfy all the Maxwell field equations
with T.E. source, Eq.(\ref{eq:rectilinear T.E. source}), whenever $\psi$
satisfies the inhomogeneous scalar wave equation, Eq.(\ref{eq:the full equation}).

\subsubsection{The T.M. Field}
\label{The T.M. Field}

For the T.M. degrees of freedom the source and the electromagnetic
field are also derived from a solution to the same inhomogeneous
scalar wave Eq.(\ref{eq:the full equation}). However, the difference
from the T.E. case is that the four-vector components of the source and
the vector potential lie in the Lorentz $(t,z)$-plane.  Thus, instead
of Eqs.(\ref{eq:rectilinear T.E. source}) and (\ref{eq:rectilinear
T.E. vector potential}), one has the T.M. source
\begin{equation}
( S_{t}, S_{z}, S_{x}, S_{y})=\left(
\frac{\partial S}{\partial z},
\frac{\partial S}{\partial t},0,0 \right)
\label{eq:rectilinear T.M. source}
\end{equation}
and the T.M. vector potential
\begin{equation}
( A_{t}, A_{z}, A_{x}, A_{y})=\left(
\frac{\partial \psi}{\partial z},
\frac{\partial \psi}{\partial t} ,0,0 \right)~.
\label{eq:rectilinear T.M. vector potential}
\end{equation}
All the corresponding T.M. field components are derived from the 
scalar $\psi(t,z,x,y)$:
\begin{equation}
\begin{array}{c}
\begin{tabular}[t]{|l|c|}	
\hline
$ E_{long.}:$
& $ \displaystyle F_{zt}= \left( 
\frac{\partial^2}{\partial z^2}
 -\frac{\partial^2}{\partial t^2} \right) \psi	$\\ 
								[1mm]\hline
$ E_x:$&$ \displaystyle  F_{xt}=
 \frac{\partial }{\partial x}\frac{\partial \psi}{\partial z } $\\
								[3mm]	\hline
$ E_y:$  &  $\displaystyle F_{yt}=
\frac{\partial }{\partial y}\frac{\partial \psi}{\partial z} $
								\\[3mm]\hline
$ B_{long.}:$ & $\displaystyle F_{xy}=0 $\\	[2mm]\hline
$ B_x:$ & $\displaystyle F_{yz}=
\frac{\partial }{\partial y}
 \frac{\partial \psi }{\partial t}	$\\
								[3mm]\hline
$ B_y:$ & $\displaystyle F_{zx}=
-\frac{\partial }{\partial x}
		\frac{\partial \psi}{\partial t}$\\
								[3mm]\hline
\end{tabular}
\end{array}
\end{equation}
These components are guaranteed to satisfy all the Maxwell field equations
with the T.M. source, Eq.(\ref{eq:rectilinear T.M. source}), whenever $\psi$
satisfies the inhomogeneous scalar wave equation, Eq.(\ref{eq:the full equation}).

\subsubsection{The T.E.M. Field Equations}
There are also the T.E.M. degrees of freedom.
For them the Maxwell four-vector source
\begin{equation}
( S_{t}, S_{z}, S_{x}, S_{y})=\left( \frac{\partial I}{\partial t},
\frac{\partial I}{\partial z},\frac{\partial J}{\partial
x},\frac{\partial J}{\partial y}\right)
\label{eq:rectilinear T.E.M. source}
\end{equation}
is derived from two functions $I(t,z,x,y)$ and, $J(t,z,x,y)$, scalars
on the 2-D Lorentz plane and the 2-D-Euclidean plane respectively. They
are, however, not independent.  Charge conservation demands the
relation
\[
	\left(
{\partial^2  \over \partial t^2}
-{\partial^2  \over \partial z^2}
	\right)I =
	\left(
{\partial^2  \over \partial x^2}+
{\partial^2  \over \partial y^2}
	\right)J
\]
The T.E.M. four-vector potential
\begin{equation}
( A_{t}, A_{z}, A_{x}, A_{y})=\left( \frac{\partial \phi}{\partial t},
\frac{\partial \phi}{\partial z},\frac{\partial \psi}{\partial
x},\frac{\partial \psi}{\partial y}\right)
\label{eq:rectilinear T.E.M. vector potential}
\end{equation}
has the same form, but only the difference $\phi-\psi$ is determined
by the field equations. Indeed, the T.E.M. field components are derived from 
this difference:
%
%
%
\begin{center}
\begin{tabular}[t]{|l|c|}	
\hline
$ E_{long.}:$
& $ \displaystyle F_{zt}= 0$\\ 
								[1mm]\hline
$ E_x:$&$ \displaystyle  F_{xt}=\partial_x A_t-\partial_t A_x=
 \frac{\partial }{\partial x}\frac{\partial (\phi-\psi)}{\partial t} $\\
								[3mm]	\hline
$ E_y:$  &  $\displaystyle F_{yt}=\partial_y A_t-\partial_t A_y=
\frac{\partial }{\partial y}\frac{\partial (\phi-\psi)}{\partial t} $
								\\[3mm]\hline
$ B_{long.}:$ & $\displaystyle F_{xy}=0 $\\	[2mm]\hline
$ B_x:$ & $\displaystyle F_{yz}=\partial_y A_z-\partial_z A_y=
\frac{\partial }{\partial y}
 \frac{\partial (\phi-\psi) }{\partial z}	$\\
								[3mm]\hline
$ B_y:$ & $\displaystyle F_{zx}=\partial_z A_x-\partial_x A_z=
-\frac{\partial }{\partial x}
		\frac{\partial (\phi-\psi)}{\partial z}$\\
								[3mm]\hline
\end{tabular}
\end{center}
This e.m. field satisfies the Maxwell field equations if any two of
the following three scalar equations,
\begin{eqnarray}
	-\left(
{\partial^2  \over \partial x^2}+
{\partial^2  \over \partial y^2}
	\right)(\phi-\psi)&=&4\pi I \\
\left(
-{\partial^2  \over \partial t^2}+
{\partial^2  \over \partial z^2}
	\right)(\phi-\psi)&=&4\pi J  \\
\left(
{\partial^2  \over \partial t^2}
-{\partial^2  \over \partial z^2}
	\right)I &=&
	\left(
{\partial^2  \over \partial x^2}+
{\partial^2  \over \partial y^2}
	\right)J
\label{eq:three T.E.M. equations}
\end{eqnarray}
are satisfied. The last equation is, of course, simply the
conservation of charge equation. Furthermore, it is evident that the T.E.M. field
propagates strictly along the $z$-axis, the direction of the Pointing
vector.

\subsection{Existence and Uniqueness of the Method of the 2+2 Split}

Can an arbitrary vector potential be written in terms of four suitably
chosen scalars $\psi^{T.E.},\psi^{T.M.},\psi$ and $\phi$ so as to
satisfy Eqs.(\ref{eq:rectilinear T.E. vector potential}),
(\ref{eq:rectilinear T.M. vector potential}), and (\ref{eq:rectilinear
T.E.M. vector potential})? If the answer is ``yes'' then any e.m. field
can be expressed in terms of these scalars, and one can claim that
these scalars give a complete and equivalent description of the
e.m. field.  It turns out that this is indeed the case. In fact, the
description is also unique.  Indeed, given the vector potential,
Eq.(\ref{eq:the vector potential}), there exist four unique scalars
which are determined by this vector potential so as to satisfy
Eqs.(\ref{eq:rectilinear T.E. vector potential}), (\ref{eq:rectilinear
T.M. vector potential}), and (\ref{eq:rectilinear T.E.M. vector
potential}). The determining equations, obtained by taking suitable
derivatives, are
\begin{eqnarray}
\frac{\partial^2 \psi^{T.E.}}{\partial x^2}+
\frac{\partial^2 \psi^{T.E.}}{\partial y^2}
&=&\partial_y A_x -\partial_x A_y \label{eq:1st equation}\\
-\frac{\partial^2 \psi^{T.M.}}{\partial t^2}+
\frac{\partial^2 \psi^{T.M.}}{\partial z^2}
&=&\partial_z A_t -\partial_t A_z \label{eq:2nd equation}\\
\frac{\partial^2 \psi}{\partial x^2}+
\frac{\partial^2 \psi}{\partial y^2}
&=&\partial_x A_x +\partial_y A_y \label{eq:3rd equation}\\ 
-\frac{\partial^2 \phi}{\partial t^2}+
\frac{\partial^2 \phi}{\partial z^2}
&=&-\partial_t A_t +\partial_z A_z \label{eq:4th equation}
\end{eqnarray}
These equations guarantee the existence of the sought after scalar
functions $\psi^{T.E.},\psi^{T.M.},\psi$ and $\phi$ . Their uniqueness
follows from their boundary conditions in the Euclidean $(x,y)$-plane
and their initial conditions in the Lorentzian
$(t,z)$-plane. Consequently, Eqs.(\ref{eq:1st equation})-(\ref{eq:4th
equation}) together with Eqs.(\ref{eq:rectilinear T.E. vector
potential}), (\ref{eq:rectilinear T.M. vector potential}), and
(\ref{eq:rectilinear T.E.M. vector potential}) establish a one-to-one
linear correspondence between the space of vector potentials and the
space of four ordered scalars,
\[
(\psi^{T.E.},\psi^{T.M.},\psi,\phi)\leftarrow \!\!\!\rightarrow
(A_t,A_z,A_x,A_y)
\]
Of the four scalars, three are gauge invariants, namely
$\psi^{T.E.}$, $\psi^{T.M.}$, and the difference $\psi -\phi$, a
result made obvious by inspecting Eqs.(\ref{eq:1st
equation})-(\ref{eq:4th equation}).

\subsection{Historical Remarks}

The T.E. scalar and the T.M. scalar whose derivatives yield the
respective vector potentials Eqs.(\ref{eq:rectilinear T.E. vector
potential}) and (\ref{eq:rectilinear T.M. vector potential}) can be
related to Righi's magnetic ``super potential'' vector $\vec{\Pi}^m$
and Hertz's electric ``super potential'' vector
$\vec{\Pi}^e$\cite{Phillips}. Indeed, if $\psi_{T.E.}$ is the
T.E. scalar and $\psi_{T.M.}$ is the T.M. scalar, then these scalars are
simply the $z$-components of the corresponding super potential vectors
\begin{eqnarray}
\vec{\Pi}^m &\equiv& (\Pi^m_x ,  
		\Pi^m_y,\Pi^m_z)=(0,0,\psi_{T.E.})~~~\textrm{``Righi''}\\
\vec{\Pi}^e &\equiv& (\Pi^e_x, 
		\Pi^e_y,\Pi^e_z)=(0,0,\psi_{T.M.})~~~~~\textrm{``Hertz''} 
\end{eqnarray}
In fact, subsequent to Hertz's 1900 and Righi's 1901 introduction of
their super potential vectors, Whittaker in 1903 showed that the
Maxwell field can be derived precisely from our two gauge invariant
scalars $\psi^{TE}$ and $\psi^{TM}$\cite{Phillips}.

\subsection{Application to Accelerated and Expanding Inertial Frames}

A key virtue of splitting spacetime according to the 2+2 scheme is its
flexibility.  It accommodates the necessary Rindler coordinate
geometries which are called for by the physical problem: accelerated
frames for the accelerated sources, and expanding inertial frames for
the inertial observers who measure the radiation emitted from these
sources. These geometries are
\begin{eqnarray}
ds^2&=&-\xi^2 d\tau^2 +d\xi^2+dr^2 +r^2d\theta^2 \qquad
\textrm{in}~I~\textrm{or~in}~II \quad \textrm{(``accelerated
~frame'')}
\label{eq:Rindler metric I and II}
\end{eqnarray}
and
\begin{eqnarray}
ds^2&=&-d\xi^2 +\xi^2 d\tau^2 +dr^2 +r^2d\theta^2 \qquad \textrm{in}~F~\textrm{or in}~P 
\quad\textrm{(``expanding (or contracting) inertial frame'')}
\label{eq:Rindler metric P and F}
\end{eqnarray}
In these two frames the Rindler/polar-coordinatized version of
Eq.(\ref{eq:the full equation}) is 
\begin{equation}
\left[
\left( 
-\frac{1}{\xi^2} \frac{\partial ^2}{\partial \tau} + 
\frac{1}{\xi} \frac{\partial}{\partial \xi}\xi \frac{\partial}{\partial \xi}
\right)+ 
\left(
\frac{1}{r} \frac{\partial}{\partial r}r \frac{\partial}{\partial r}
+\frac{1}{r^2} \frac{\partial ^2}{\partial \theta}
\right)
\right]
\psi(\tau,\xi,r,\theta)=-4\pi S(\tau,\xi,r,\theta)~~~~~~~~~~~~
		\textrm{in}~I~\textrm{or~in}~II~,
\label{eq:full scalar equation in I and II}
\end{equation}
and 
\begin{equation}
\left[
\left( 
-\frac{1}{\xi} \frac{\partial}{\partial \xi}\xi \frac{\partial}{\partial \xi}
+\frac{1}{\xi^2} \frac{\partial ^2}{\partial \tau}
\right)+ 
\left(
\frac{1}{r} \frac{\partial}{\partial r}r \frac{\partial}{\partial r}
+\frac{1}{r^2} \frac{\partial ^2}{\partial \theta}
\right)
\right]
\psi(\xi,\tau, r,\theta)=0~~~~~~~~~~~~~~~~~~~~~
		\textrm{in}~F~\textrm{or~in}~P~.
\label{eq:full scalar equation in F and P}
\end{equation}
\emph{Notational rule:} The Rindler coordinates listed in the
arguments of the scalar functions in Eqs.(\ref{eq:full scalar equation
in I and II}) and (\ref{eq:full scalar equation in F and P}) are
always listed with the timelike coordinate first, followed by the
spatial coordinates. Thus $(\tau,\xi,r,\theta)$ implies that the
function is defined on Rindler sectors $I$ or $II$, as in
Eq.(\ref{eq:full scalar equation in I and II}). On the other hand,
$(\xi,\tau,r,\theta)$ implies that the domain of the function is
$F$ or $P$, as in Eq.(\ref{eq:full scalar equation in F and P}).

The feature common to the T.E. and the T.M. field is that both 
of them are based on the two-dimensional curl of a scalar, say $\psi$. 
The difference is that for the T.E. field this curl is in the Euclidean plane,
\[
\nabla_a \times \psi \equiv \epsilon_{ab} ~^{(2)} g^{bc}
\frac{\partial \psi}{\partial x^c}:~~(\nabla_r \times \psi,\nabla_\theta 
\times \psi)= \left( \frac{1}{r} \frac{\partial \psi}{\partial\theta},
-r \frac{\partial \psi}{\partial r} \right)~,
\]
while for the T.M. field this curl is the Lorentz plane,
\[
\nabla_A \times \psi \equiv \epsilon_{AB} ~^{(2)} g^{BC}
\frac{\partial \psi}{\partial x^C}:~~(\nabla_\tau \times \psi,\nabla_\xi 
\times \psi)= \left( \xi \frac{\partial \psi}{\partial \xi},
\frac{1}{\xi} \frac{\partial \psi}{\partial\tau} \right)\quad \textrm{in }I
\textrm{ or in }II~,
\]
and 
\[
\nabla_A \times \psi \equiv \epsilon_{AB} ~^{(2)} g^{BC}
\frac{\partial \psi}{\partial x^C}:~~(\nabla_\xi \times \psi,\nabla_\tau 
\times \psi)= \left( \frac{1}{\xi} \frac{\partial \psi}{\partial\tau},
\xi \frac{\partial \psi}{\partial \xi} \right)\quad \textrm{in }F
\textrm{ or in }P~,
\]
The $\epsilon_{ab}$ and $\epsilon_{AB}$ are the components of the
antisymmetric area tensors on the two respective planes.

\subsubsection{The T.E. Field}
\label{sec-the T.E. field}
For the T.E. degrees of freedom the charge-current $S_\mu dx^\mu$ and
the vector potential $A_\mu dx^\mu$ have the form given by 
\begin{equation}
( S_{\tau}, S_{\xi}, S_{r}, S_{\theta})=\left(0,0,
\frac{1}{r}\frac{\partial S}{\partial \theta},
-r\frac{\partial S}{\partial r}\right)~,
\label{eq:polar T.E. source}
\end{equation}
and
\begin{equation}
( A_{\tau}, A_{\xi}, A_{r}, A_{\theta})=\left(0,0,
\frac{1}{r}\frac{\partial \psi}{\partial \theta},
-r\frac{\partial \psi}{\partial r}\right)~.
\label{eq:polar T.E. vector potential}
\end{equation}
The electromagnetic field,
\begin{eqnarray*}
\frac{1}{2}F_{\mu\nu}dx^\mu \wedge dx^\nu &\equiv&
\hat E_{long.}d\xi\wedge \xi d\tau +\hat E_r dr\wedge \xi d\tau 
				+\hat E_\theta r d\theta \wedge 	\xi d\tau\\
&~&~+\hat B_r rd\theta\wedge d\xi +\hat B_\theta d\xi\wedge dr +\hat B_{long.}dr\wedge rd\theta
		\qquad\qquad \qquad\qquad\textrm{in}~I~\textrm{and}~II~\\
&\equiv&\hat E_{long.}\xi d\tau\wedge d\xi +\hat E_r dr\wedge d\xi 
					+\hat E_\theta rd\theta\wedge d\xi \\
&~&~+\hat B_r rd\theta\wedge \xi d\tau +\hat B_\theta \xi d\tau\wedge dr +\hat B_{long.}dr\wedge rd\theta 
		\qquad\qquad \qquad\quad\textrm{in}~P~\textrm{and}~F~,
\end{eqnarray*}
has the following components:
\begin{center}
\begin{tabular}[t]{|l|c|c|}	
\hline			
~		&In $I$ or in $II$		&In $F$ or in $P$\\ \hline
$\hat E_{long.}:$& $  \displaystyle{\frac{1}{\xi}F_{\xi\tau}=0}$ 
			& $\displaystyle  \frac{1}{\xi}F_{\tau\xi}=0$\\ [3mm]\hline

$\hat E_r:$&$   \displaystyle\frac{1}{\xi}F_{r\tau}=
	-\frac{1}{r} \frac{\partial}{\partial\theta}\left( \frac{1}{\xi}
					\frac{\partial \psi}{\partial \tau} \right)$
			&$ \displaystyle  F_{r\xi}=
-\frac{1}{r} \frac{\partial}{\partial\theta}\left( 
					\frac{\partial \psi}{\partial \xi} \right) $\\
								[3mm]	\hline
$\hat E_\theta:$& $  \displaystyle\frac{1}{\xi r}F_{\theta\tau}=
  \frac{\partial}{\partial r}\left( \frac{1}{\xi}\frac{\partial \psi}{\partial \tau}
									\right)$
			& $\displaystyle  \frac{1}{r}F_{\theta\xi}=
\frac{\partial}{\partial r}\left( \frac{\partial \psi}{\partial \xi} \right)$\\
									[3mm]\hline
$\hat B_{long.}:$&$   \displaystyle \frac{1}{r}F_{r\theta}=
-\left( 
\frac{1}{r}\frac{\partial}{\partial r}r \frac{\partial}{\partial r}
 +\frac{1}{r^2}\frac{\partial^2}{\partial\theta^2} \right) \psi
$
	& $\displaystyle \frac{1}{r}  F_{r\theta}=
-\left( 
\frac{1}{r}\frac{\partial}{\partial r}r \frac{\partial}{\partial r}
 +\frac{1}{r^2}\frac{\partial^2}{\partial\theta^2} \right) \psi
$\\
									[2mm]\hline
$\hat B_r:$&$   \displaystyle \frac{1}{r}F_{\theta\xi}=
  \frac{\partial}{\partial r}\left(\frac{\partial \psi}{\partial \xi}\right)$
			& $\displaystyle  \frac{1}{\xi r}F_{\theta\tau }=
\frac{\partial}{\partial r}\left( \frac{1}{\xi}\frac{\partial \psi}{\partial \tau}
							\right) $\\[3mm]\hline
$\hat B_\theta:$&$   \displaystyle F_{\xi r}=
 \frac{1}{r}\frac{\partial}{\partial\theta} \left(\frac{\partial \psi}{\partial \xi}
									\right)$
			& $\displaystyle \frac{1}{\xi}F_{\tau r}=
\frac{1}{r}\frac{\partial}{\partial\theta} \left( 
		\frac{1}{\xi}\frac{\partial \psi}{\partial \tau}	\right)
								$\\[3mm] \hline
\end{tabular}
\end{center}
The carets in the first column serve as a reminder that these
components are relative to the orthonormal basis of the metric,
Eqs.(\ref{eq:Rindler metric I and II}) and (\ref{eq:Rindler metric P and F}).

\subsubsection{The T.M. Field}
The T.M. has its source and vector potential four-vectors lie strictly
in the 2-d Lorentz plane:
\begin{equation}
( S_{\tau}, S_{\xi}, S_{r}, S_{\theta})=\left(
\xi\frac{\partial S}{\partial \xi},
\frac{1}{\xi}\frac{\partial S}{\partial \tau},0,0 \right)\quad
				\textrm{in }I\textrm{ or in }II~,
\label{eq:polar T.M. source}
\end{equation}
\begin{equation}
( A_{\tau}, A_{\xi}, A_{r}, A_{\theta})=\left(
\xi\frac{\partial \psi}{\partial \xi},
\frac{1}{\xi}\frac{\partial \psi}{\partial \tau},0,0\right)\quad
				\textrm{in }I\textrm{ or in }II~,
\label{eq:polar T.M. vector potential in I or II}
\end{equation}
and
\begin{equation}
( A_{\xi}, A_{\tau}, A_{r}, A_{\theta})=\left(
\frac{1}{\xi}\frac{\partial \psi}{\partial \tau},
			\xi\frac{\partial \psi}{\partial \xi},0,0\right)\quad
				\textrm{in }F\textrm{ or in }P~
\label{eq:polar T.M. vector potential in F or P}
\end{equation}
The components of the T.M. Maxwell field are
\begin{center}
\begin{tabular}[t]{|l|c|c|}	
\hline			
~		&In $I$ or in $II$		&In $F$ or in $P$\\ \hline
$\hat E_{long.}:$& $  \displaystyle \frac{1}{\xi}F_{\xi\tau}=
\left( 
\frac{1}{\xi}\frac{\partial}{\partial \xi}\xi\frac{\partial}{\partial \xi}
 -\frac{1}{\xi^2}\frac{\partial^2}{\partial\tau^2} \right) \psi$ 
			& $\displaystyle  \frac{1}{\xi}F_{\tau\xi}=
-\left( 
\frac{1}{\xi}\frac{\partial}{\partial \xi}\xi\frac{\partial}{\partial \xi}
 -\frac{1}{\xi^2}\frac{\partial^2}{\partial\tau^2} \right) \psi $\\ [3mm]\hline

$\hat E_r:$&$   \displaystyle\frac{1}{\xi}F_{r\tau}=
	\frac{\partial}{\partial r}\left(
					\frac{\partial \psi}{\partial \xi} \right)$
			&$ \displaystyle  F_{r\xi}=
\frac{\partial}{\partial r}\left( 
			\frac{1}{\xi}\frac{\partial \psi}{\partial \tau} \right) $\\
								[3mm]	\hline
$\hat E_\theta:$& $  \displaystyle\frac{1}{\xi r}F_{\theta\tau}=
 \frac{1}{r}  \frac{\partial}{\partial \theta}
				\left( \frac{\partial \psi}{\partial \xi}
									\right)$
			& $\displaystyle  \frac{1}{r}F_{\theta\xi}=
\frac{1}{r} \frac{\partial}{\partial \theta}\left
		( \frac{1}{\xi}\frac{\partial \psi}{\partial \tau} \right)$\\
									[3mm]\hline
$\hat B_{long.}:$&$   \displaystyle \frac{1}{r}F_{r\theta}=0$
	& $\displaystyle \frac{1}{r}  F_{r\theta}=0$\\
									[2mm]\hline
$\hat B_r:$&$   \displaystyle \frac{1}{r}F_{\theta\xi}=
\frac{1}{r}\frac{\partial}{\partial \theta}
		\left(\frac{1}{\xi}\frac{\partial \psi}{\partial \tau}\right)$
			& $\displaystyle  \frac{1}{\xi r}F_{\theta\tau }=
\frac{1}{r}\frac{\partial}{\partial \theta}\left( 
			\frac{\partial \psi}{\partial \xi}
							\right) $\\[3mm]\hline
$\hat B_\theta:$&$   \displaystyle F_{\xi r}=
 -\frac{\partial}{\partial r} \left( \frac{1}{\xi}\frac{\partial \psi}{\partial \tau}
									\right)$
			& $\displaystyle \frac{1}{\xi}F_{\tau r}=
-\frac{\partial}{\partial r} \left( 
		\frac{\partial \psi}{\partial \xi}	\right)
								$\\[3mm] \hline
\end{tabular}
\end{center}

\subsubsection{A Mnemonic Short Cut}
There is a quick way of obtaining all the physical (orthonormal)
components of the electric and magnetic field.  Note that the
longitudinal electric and magnetic field components $\hat E_{long}$ and
$\hat B_{long}$ are scalars in the Lorentz plane and in the Euclidean
plane transverse to it. Consequently, for these components the
transition from Minkowski to Rindler/polar coordinates could have been
done without any computations. The same is true for the
two-dimensional transverse electric and magnetic field vectors.  As
suggested by Eqs.(\ref{eq:Rindler metric I and II}) and
(\ref{eq:Rindler metric P and F}), in the denominator of the partial
derivatives simply make the replacements
\begin{eqnarray*}
\partial t &\rightarrow& \xi \partial \tau\\
\partial z &\rightarrow& \partial \xi\\
\partial x &\rightarrow& \partial r\\
\partial y &\rightarrow& r \partial \theta
\end{eqnarray*}
in Rindler sectors $I$ or $II$, and  
\begin{eqnarray*}
\partial t &\rightarrow& \partial \xi\\
\partial z &\rightarrow& \xi \partial \tau\\
\partial x &\rightarrow& \partial r\\
\partial y &\rightarrow& r \partial \theta
\end{eqnarray*}
in Rindler sectors $F$ or $P$. These replacements yield the computed 
transverse T.E. and T.M. components. 

There also is a quick way of obtaining the T.M. field from the
T.E. field components. Let $\psi^{TE}$ be the scalar wave function
which satisfies the Klein-Gordon wave function for the T.E. field,
and let $\psi^{TM}$ be that for the T.M. field. Then the corresponding 
field components are related as follows:
\begin{eqnarray*}
T.E. &~& ~T.M.\\
\psi^{TE}	&\rightarrow&	~~\psi^{TM}	\\
\hat E_{long}=0	&\rightarrow& -\hat B_{long}=0	\\
\hat E_r	&\rightarrow& -\hat B_r		\\
\hat E_\theta	&\rightarrow& -\hat B_\theta	\\
\hat B_{long}	&\rightarrow& ~~\hat E_{long}~~~~~~~(in~vacuum)\\
\hat B_r	&\rightarrow& ~~\hat E_r		\\
\hat B_\theta	&\rightarrow& ~~\hat E_\theta	
\end{eqnarray*}	
This relationship holds in all four Rindler sectors. It also holds
correspondingly relative to the rectilinear coordinate frame in
Sections \ref{The T.M. Field} and \ref{The T.E. Field}.

\section{RADIATION: MATHEMATICAL RELATION TO THE SOURCE}
\label{sec-radiation:mathematical}

Any Maxwell field $F_{\mu\nu}$ can be obtained from a single
Klein-Gordon scalar field $\psi$, a solution to the scalar wave
Eq.(\ref{eq:the full equation}). This is done with the help of the
T.E. and T.M. tables of derivatives. Similarly, any K-G field $\psi$
can be obtained from the source function $S$. This is done with the
help of the unit impulse response $\mathcal G$ (Green's function),
the solution to
\begin{equation}
\left( 
-\frac{\partial ^2}{\partial t^2} + 
\frac{\partial^2}{\partial z^2}+ 
\frac{1}{r} \frac{\partial}{\partial r}r \frac{\partial}{\partial r}
+\frac{1}{r^2} \frac{\partial ^2}{\partial \theta}~.
\right)
{\mathcal{G}}
=-\delta(t-t')\delta(z-z') \frac{\delta(r-r')}{r}
\delta(\theta-\theta') 
\label{eq:equation for the unit impulse response}
\end{equation}
In terms of $\mathcal G$ the solution to the inhomogeneous wave
equation, Eq.(\ref{eq:full scalar equation in I and II}) is 
\begin{eqnarray}
\psi(t,z,r,\theta)&=&
\int_{-\infty}^\infty \int_0^\infty \int_0^\infty \int_0^{2\pi}
{\mathcal{G}}(t,z,r,\theta;\tau',\xi',r',\theta')
~4\pi S(t',z',r',\theta')~~dt'\, dz'\, r' dr'\, d\theta'~,
\label{eq:full scalar fields in I or II}
\end{eqnarray}
Here $S$ is the scalar source, which is non-zero only in Rindler
sectors $I$ and $II$.

\subsection{Unit Impulse Response}
The solution to Eq.(\ref{eq:equation for the unit impulse response})
is the retarded Green's function, a unique scalar field, whose domain
extends over all four Rindler sectors. One accommodates the cylindrical
symmetry of the coordinate geometry by representing the scalar
field in terms of the appropriate eigenfunctions, the Bessel harmonics
$J_m(kr)e^{im\theta}$, for the Euclidean $(r,\theta)$-plane:
\begin{equation}
{\mathcal{G}}(t,z,r,\theta;t',z',r',\theta')=
\sum_{m=-\infty}^\infty \int_0^\infty
G(t,z;t',z') ~J_m(kr)\frac{e^{im(\theta-\theta')}}{2\pi}
                        J_m(kr')~k\, dk ~,
\label{eq:global Green's function in terms of Bessel harmonics}
\end{equation}
where $G$ satisfies
\begin{equation}
\left(
-\frac{\partial ^2}{\partial t^2}+
\frac{\partial^2}{\partial z^2} - k^2
\right)
G=-\delta(t-t')\delta(z-z')~,
\end{equation}
and
\[
G=0 \quad \textrm{whenever }t<t'~.
\]
This Green's function is unique, and it is easy to show\cite{HOW TO FIND IT} that 
\begin{equation}
G=\left\{
	\begin{array}{ll}
	\frac{1}{2}J_0(k\sqrt{(t-t')^2-(z-z')^2}) 
					&\textrm{ whenever }t-t'\ge\vert z-z'\vert\\
	0&	\textrm{ whenever }			t-t'<\vert z-z'\vert~.
	\end{array}
  \right.
\label{eq:partial global Green's function}
\end{equation}
This means that $G$ is non-zero only inside the future of the source
event $(t',z')$, and vanishes identically everywhere else.  The
function $G(t,z;t',z')$ is defined on all four Rindler
sectors. However, our interest is only in those of its coordinate
representatives whose source events lie Rindler sectors $I$ or $II$,
\[
\left.
\begin{array}{ll}
t'&=\pm\xi'\sinh\tau'\\
z'&=\pm\xi'\cosh\tau'
\end{array}
\right\}
\left\{
\begin{array}{l}
\textrm{ upper sign for }I\\
\textrm{ lower sign for }II~,
\end{array}
\right.
\]
and whose observation events lie in Rindler sector $F$,
\[
\begin{array}{ll}
t&=\xi\cosh\tau\\
z&=\xi\sinh\tau ~.
\end{array}
\]
For these coordinate restrictions the two coordinate representatives
of $G(t,z;t',z')$, Eq.(\ref{eq:partial global Green's function}), are
\begin{eqnarray}
G_I(k\xi,\tau;\tau',k\xi')&=&
\left\{
	\begin{array}{ll}
	\frac{1}{2}
		J_0\left(k\sqrt{\xi^2-\xi'^2+ 2\xi\xi' \sinh (\tau-\tau')}\right)
          &\textrm{whenever}~(\xi,\tau)~\textrm{is~in}~F~ 
\textrm{and}~(\xi',\tau') ~\textrm{is~in}~I \\
	0 &\textrm{  whenever }\xi^2-\xi'^2+2\xi\xi'\sinh(\tau-\tau')<0
	\end{array}
\right.
		\label{eq:partial Green's function for sector I} 
\end{eqnarray}
and
\begin{eqnarray}
G_{II}(k\xi,\tau;\tau',k\xi')&=&
\left\{
	\begin{array}{ll}
	\frac{1}{2}
		J_0\left(k\sqrt{\xi^2-\xi'^2-2\xi\xi' \sinh (\tau-\tau')}\right)
          &\textrm{whenever}~(\xi,\tau)~\textrm{is~in}~F~ 
\textrm{and}~(\xi',\tau') ~\textrm{is~in}~II \\
	0 &\textrm{  whenever }\xi^2-\xi'^2-2\xi\xi'\sinh(\tau-\tau')<0
	\end{array}
\right.
		\label{eq:partial Green's function for sector II}
\end{eqnarray}
These two coordinate representatives give rise to the corresponding
two representatives of the unit impulse response, Eq.(\ref{eq:global
Green's function in terms of Bessel harmonics}),
\begin{equation}
{\mathcal{G}}_{I,II}(\xi,\tau,r,\theta;\tau',\xi',r',\theta')=\int_0^\infty
G_{I,II}(k\xi,\tau;\tau',k\xi') \sum_{m=-\infty}^\infty 
J_m(kr)\frac{e^{im(\theta-\theta')}}{2\pi}
                        J_m(kr')~k\, dk ~,
\label{eq:Green's function representatives in terms of Bessel harmonics}
\end{equation}
This integral expression is exactly what is needed to obtain the
radiation field from bodies accelerated in $I$ and/or $II$. However,
in order to ascertain agreement with previously established knowledge,
we shall use the remainder of this subsection to evaluate the sum and
the integral in Eq.(\ref{eq:Green's function representatives in terms
of Bessel harmonics}) explicitly.

It is a delightful property of Bessel harmonics that the sum over $m$
can be evaluated in closed form\cite{Sommerfeld}. This property is the Euclidean plane
analogue of what for spherical harmonics is the spherical addition
theorem.  One has
\begin{equation}
\sum_{m=-\infty}^\infty 
J_m(kr)\frac{e^{im(\theta-\theta')}}{2\pi} J_m(kr')=\frac{1}{2\pi}
J_0\left(k\sqrt{r^2+r'^2 -2rr'\cos(\theta-\theta')}\right)
\label{eq:addition theorem for the Euclidean plane}
\end{equation}
Inserting this result, as well as Eqs.(\ref{eq:partial Green's function
for sector I}) or (\ref{eq:partial Green's function for sector II})
into Eq.(\ref{eq:Green's function representatives in terms of Bessel
harmonics}) yields the two unit impulse response functions with
sources in $I$ (upper sign) and $II$ (lower sign) 
\begin{eqnarray}
{\mathcal{G}}_{I,II}(\xi,\tau,r,\theta;\tau',\xi',r',\theta')
&=& \frac{1}{4\pi}\int_0^\infty 
J_0\left(k\sqrt{\xi^2-\xi'^2\pm 2\xi\xi' \sinh (\tau-\tau')}\right)
J_0\left(k\sqrt{r^2+r'^2 -2rr'\cos(\theta-\theta')}\right) k\,dk
							\nonumber \\
&=& \frac{1}{4\pi}\int_0^\infty
J_0\left( k\sqrt{(t-t')^2-(z-z')^2}\right)J_0\left(k\sqrt{(x-x')^2-(y-y')^2}\right) 
 \,k\,dk\nonumber 
\end{eqnarray}
whenever $t-t' \ge \vert z-z'\vert$ and zero otherwise.
The spread-out amplitudes of this linear superposition interfere
constructively to form a Dirac delta function response. Indeed, 
using the standard representation
\[
\int_0^\infty J_0(ka)J_0(kb)~kdk=\frac{\delta(a-b)}{b}
\]
for this function, one finds that
\begin{eqnarray}
{\mathcal{G}}_{I,II}(\xi,\tau,r,\theta;\tau',\xi',r',\theta')&=&\frac{1}{4\pi}
\frac{\delta\left(
\sqrt{\xi^2-\xi'^2\pm 2\xi\xi' \sinh (\tau-\tau')}-
\sqrt{r^2+r'^2 -2rr'\cos(\theta-\theta')} 
	\right)}{\sqrt{r^2+r'^2 -2rr'\cos(\theta-\theta')}}\nonumber\\
&=&\frac{1}{2\pi}
\delta\left(
\xi^2-\xi'^2\pm 2\xi\xi' \sinh (\tau-\tau')- (r^2+r'^2 -2rr'\cos(\theta-\theta'))
	\right)\label{causal response b}\\
&=&\frac{1}{2\pi}
\delta\left( (t-t')^2-(z-z')^2-(x-x')^2-(y-y')^2 \right)
\nonumber
\end{eqnarray}
whenever $(t,z,x,y)$ is in the future of $(t',z',x',y')$.
This is the familiar causal response in $F$ due to a unit impulse event in 
$I$ or in $II$. 

\subsection{Full Scalar Radiation Field}

The scalar field measured in Rindler sector $F$ of the expanding
inertial reference frame is the linear superposition
\begin{equation}
\psi_F(\xi,\tau,r,\theta)=\psi_I(\xi,\tau,r,\theta)+\psi_{II}(\xi,\tau,r,\theta)
\label{eq:full scalar field in F from I and II}
\end{equation}
of two contributions. They arise from two causally disjoint sources,
one in Rindler sector $I$, the other in Rindler sector $II$.  Applying
Eq.(\ref{eq:Green's function representatives in terms of Bessel
harmonics}) together with Eqs.(\ref{eq:partial Green's function for
sector I}) and (\ref{eq:partial Green's function for sector II}) to
Eq.(\ref{eq:full scalar fields in I or II}), one finds that these
contributions are
\begin{eqnarray}
\lefteqn{ \psi_{I,II}(\xi,\tau,r,\theta)= \sum_{m=-\infty}^\infty \int_0^\infty k\, dk e^{im\theta} J_m(kr) \times }\nonumber\\
& &\int_{-\infty}^\infty d\tau' \int_0^\infty \xi' d\xi' \frac{1}{2}
J_0\left( k\sqrt{\xi^2-\xi'^2 \pm 2\xi\xi' \sinh (\tau-\tau')}\right)
\int_0^\infty r'dr' \int_0^{2\pi} d\theta' \frac{e^{-im\theta'}}{2\pi}J_m(kr')
~4\pi S_{I,II}(\tau',\xi',r',\theta')
\label{eq:two contributions from I and II}
\end{eqnarray}
These radiation fields are exact in the sense that we have made no
assumptions about the size of the sources in relation to their
radiated wavelengths. The only restriction is that the two source
functions $S_I$ and $S_{II}$ be non-zero only in $I$ and $II$
respectively.

The key to identifying the nature of these fields is the evaluation of
the mode sum over $m$ and the mode integral over $k$.
These evaluations have been done already starting with
Eq.(\ref{eq:addition theorem for the Euclidean plane}), and gave rise
to the familiar causal unit impulse response, Eq.(\ref{causal response
b}). However, for the purpose of physical transparency we shall not
make these evaluations.  Instead, we shall make an evaluation which is
based on the assumption that each source is small compared to its
radiated wave lengths $k^{-1}$. Physically this means that the phase
of the e.m. field is the same across the whole extent of the
source. Mathematically this means that the Bessel function $J_0(kr')$
should be replaced by the expression Eq.(\ref{eq:approximate Bessel
function}) below. The advantage of the long wave length approximation
is physical/technological: it permits the characterization of the
source in terms of multipole moments, which are directly tied to the
angular distribution of the emitted radiation, and which derive their
importance from, among others, the quantum mechanical selection rules.

\subsubsection{Source as a Sum of Multipoles}
The circumstance of long wave lengths is expressed by the inequality
\[
ka\ll 1 ~,
\]
where $a$ is the radius of the cylinder surrounding the source. This
circumstance allows us to set
\begin{equation}
J_m(kr')\approx \left\{ \begin{array}{ll}
			\displaystyle\frac{1}{ m !}\left( \frac{kr'}{2}
			\right)^{m } &m=0, 1,2,\cdots\\
			\displaystyle\frac{(-1)^m}{\vert m\vert !}\left( \frac{kr'}{2}
			\right)^{\vert m \vert} & m=0,-1,-2,\cdots
			\end{array} \right.
\label{eq:approximate Bessel function}
\end{equation}
throughout the integration region where the source is non-zero, and it
allows us to introduce the $(m+1)$st multipole moment (per unit length
$d\xi$)
\begin{equation}
\frac{i^m}{\vert m\vert !}
\int_0^\infty r'dr' \int_0^{2\pi} d\theta' \frac{e^{-im\theta'}}{2\pi}
\left( \frac{r'}{2}\right)^{\vert m \vert}
~4\pi S_{I,II}(\tau',\xi',r',\theta')\equiv 2 S^m_{I,II}(\tau',\xi')
\quad 
\left[\frac{\textrm{charge}}{\textrm{length}} 
			\times(\textrm{length})^{\vert m\vert +1}\right]
\label{eq:multipole}
\end{equation}
for the double integral on the right hand side of Eq.(\ref{eq:two
contributions from I and II}). This multipole density \cite{factor of
two} is complex. However, the reality of the master source
$S_{I,II}(\tau',\xi',r',\theta')$ implies and is implied by
\[
\overline{S^m_{I,II}(\tau',\xi')}=S^{-m}_{I,II}(\tau',\xi')
\]
In terms of this multipole density the full scalar radiation field in $F$ is
\begin{eqnarray}
\lefteqn{ \psi_F(\xi,\tau,r,\theta)= \sum_{m=-\infty}^\infty
\int_0^\infty dk \, k \,k^{\vert m \vert}e^{im\theta} J_m(kr) 
							\times }\nonumber\\
& &\int_{-\infty}^\infty d\tau' \int_0^\infty d\xi' \xi' \, \frac{2}{2}
\left\{ J_0\left( k\sqrt{\xi^2-\xi'^2+ 2\xi\xi' \sinh (\tau-\tau')}\right)
2S^m_{I}(\tau',\xi')+
J_0\left( k\sqrt{\xi^2-\xi'^2-2\xi\xi' \sinh (\tau-\tau')}\right)
2S^m_{II}(\tau',\xi')\right\}
\label{eq:multipole field}
\end{eqnarray}
The evaluation of the mode integral $\int_0^\infty dk\,k \cdots$is now
an easy two step task. First recall  the $m$th recursion relation
\begin{equation}
e^{im\theta} J_m(kr)= \frac{(-1)^m}{k^{\vert m\vert}} 
\left[ e^{i\theta} \left(\frac{\partial}{\partial r} +\frac{i}{r}
\frac{\partial}{\partial \theta} \right) \right]^m
J_0(kr),\quad m=0,\pm1,\pm2,\cdots
\label{eq:mth order Bessel mode}
\end{equation}
where for negative $m$ one uses
\[
\left[ e^{i\theta} \left(\frac{\partial}{\partial r} +\frac{i}{r}
\frac{\partial}{\partial \theta} \right) \right]^{-\vert m\vert} \equiv
\left[ -e^{-i\theta} \left(\frac{\partial}{\partial r} -\frac{i}{r}
\frac{\partial}{\partial \theta} \right) \right]^{\vert m\vert} ~.
\]
This recursion relation is a consequence of consolidating two familiar
contiguity relations for the Bessel functions. Introduce Eq.(\ref{eq:mth order Bessel mode}) into the integrand of Eq.(\ref{eq:multipole field}). 

Second, use the standard expression
\[
\int_0^\infty J_0(kr)J_0(k\sqrt{\cdots})k\, dk=
\frac{\delta(r-\sqrt{\cdots})}{\sqrt{\cdots}}
\]
for the Dirac delta function.  Apply this equation to
Eq.(\ref{eq:multipole field}).  Consequently, the full scalar
radiation field in Rindler sector $F$ reduces to the following
multipole expansion
\begin{eqnarray}
\psi_F(\xi,\tau,r,\theta)= \sum_{m=-\infty}^\infty 
(-1)^m \left[ e^{i\theta} \left(\frac{\partial}{\partial r} +\frac{i}{r}
\frac{\partial}{\partial \theta} \right) \right]^m \psi_m(\xi,\tau,r) \quad
\quad [\textrm{charge}]
\label{eq:multipole expansion}
\end{eqnarray}
where
\begin{eqnarray}
\psi_m(\xi,\tau,r)&=& \int_{-\infty}^\infty \int_0^\infty \frac{1}{2}
\left\{ \frac{2S^m_{I}(\tau',\xi')}{\sqrt{\xi^2-\xi'^2+2\xi\xi' \sinh
(\tau-\tau')}} \delta\left(r-\sqrt{\xi^2-\xi'^2+2\xi\xi' \sinh
(\tau-\tau')}\right)+\right.\nonumber\\
&~&~~~~~~~~~~~~~~~~\left.\frac{2S^m_{II}(\tau',\xi')}{\sqrt{\xi^2-\xi'^2-2\xi\xi'
\sinh (\tau-\tau')}} \delta\left(r-\sqrt{\xi^2-\xi'^2-2\xi\xi' \sinh
(\tau-\tau')}\right) \right\}d\tau'\xi'\,d\xi' ~.\nonumber
\end{eqnarray}
Doing the $\tau'$-integration yields
\begin{eqnarray}
\psi_m(\xi,\tau,r)
&=&
2\int_0^\infty 
\frac{\left[S^m_I(\tau',\xi')\right]_I -
\left[S^m_{II}(\tau',\xi')\right]_{II}}{\sqrt{(\xi^2-\xi'^2-r^2)^2+(2\xi\xi')^2 }}
\xi' d\xi'~.
\label{eq:multipole field with two sources}
\end{eqnarray}
Here $[~~]_I$ and $[~~]_{II}$ mean that the source functions are
evaluated in compliance with the Dirac delta functions at $\tau'=\tau
+\sinh^{-1}\frac{\xi^2-\xi'^2-r^2}{2\xi\xi'}$ and $\tau'=\tau
-\sinh^{-1}\frac{\xi^2-\xi'^2-r^2}{2\xi\xi'}$ respectively.  Recall
that $\tau'$ is a strictly timelike coordinate in Rindler sector $I$,
while in $F$ the coordinate $\tau$ is strictly spacelike. Consequently, one should not be
tempted to identify $[~~]_I$ and $[~~]_{II}$ with what in a static
inertial frame corresponds to evaluations at advanced or retarded
times. Instead, one should think of the observation event
$(\xi,\tau,r)$ in $F$ and the source event $(\tau',\xi',r )$ in $I$
as lying on each other's light cones
\[
(t-t')^2 -(z-z')^2 =r^2~,
\]
both of which cut across the future event horizons $t=\vert z\vert$ of $I$ and 
$II$. More explicitly, one has 
\[
\left[S^m_I(\tau',\xi')\right]_I\equiv S^m_I\left( \tau+\sinh^{-1}
\frac{\xi^2-\xi'^2-r^2}{2\xi\xi'},\xi' \right)~,
\]
which means that the source $S^m_I(\tau',\xi')$ has been evaluated on
the past light cone
\[
(t-t')^2-(z-z')^2\equiv\xi^2-\xi'^2+ 2\xi\xi' \sinh (\tau-\tau')=r^2
\]
of $(\xi,\tau,r)$ at $(\tau',\xi',0)$ in Rindler
sector $I$. Similarly,
\[
\left[S^m_{II}(\tau',\xi')\right]_{II}\equiv S^m_{II}\left( \tau-\sinh^{-1}
\frac{\xi^2-\xi'^2-r^2}{2\xi\xi'} ,\xi'\right)~,
\]
which means that the source $S^m_{II}(\tau',\xi')$ has been evaluated on
the past light cone 
\[
(t-t')^2-(z-z')^2\equiv\xi^2-\xi'^2- 2\xi\xi' \sinh (\tau-\tau')=r^2
\]
of $(\xi,\tau,r)$ at $(\tau',\xi',0)$ in Rindler sector $II$.  The
expression, Eq.(\ref{eq:multipole field with two sources}), for the full
scalar radiation field is exact within the context of wavelengths
large compared to the size of the source. Furthermore, one should note
that even though there is only one $\xi'$-integral, $S^m_{I}$ and
$S^m_{II}$ are source functions with distinct domains, namely, Rindler
sectors $I$ and $II$ respectively.

\subsubsection{Multipole Radiation Field}

The field is a superposition of multipole field amplitudes. The first
few terms of this superposition are
\begin{eqnarray}
\psi_F(\xi,\tau,r,\theta)&=&\psi_0(\xi,\tau,r) \nonumber \\
&-& e^{i\theta} \frac{\partial}{\partial r}\psi_1(\xi,\tau,r)\nonumber \\
&+& e^{2i\theta}\left( 
\frac{\partial^2}{\partial r^2}
-\frac{1}{r}\frac{\partial}{\partial r} 
\right)\psi_2(\xi,\tau,r)\nonumber \\
&-& e^{3i\theta}\left( 
\frac{\partial^3}{\partial r^3}
-\frac{3}{r}\frac{\partial^2}{\partial r^2}
+\frac{3}{r^2}	\frac{\partial}{\partial r} \right)\psi_3(\xi,\tau,r)\nonumber \\
&+& e^{4i\theta}\left( 
\frac{\partial}{\partial r^4}
-\frac{6}{r}\frac{\partial^3}{\partial r^3}
+\frac{12}{r^2}\frac{\partial^2}{\partial r^2}
-\frac{15}{r^3}	\frac{\partial}{\partial r} \right)\psi_4(\xi,\tau,r)~+\quad \cdots\nonumber\\
&+&(\textrm{complex~conjugate~terms}~
\textrm{corresponding~to}~m=-1,-2,-3,\cdots)
\label{eq:multipole angular distributions}
\end{eqnarray}
whose explicit form is
\begin{eqnarray}
\psi_F(\xi,\tau,r,\theta) &=&\int_0^\infty  \left[
\frac{2S^0_I(\tau+\sinh^{-1}u,\xi')-2S^0_{II}(\tau-\sinh^{-1}u,\xi')}{\sqrt{(\xi^2-\xi'^2-r^2)^2+(2\xi\xi')^2 }}
\right]\xi'\, d\xi' \nonumber \\
&-& e^{i\theta} \frac{\partial}{\partial r}
\left[
\frac{2S^1_I(\tau+\sinh^{-1}u,\xi')-2S^1_{II}(\tau-\sinh^{-1}u)}{\sqrt{(\xi^2-\xi'^2-r^2)^2+(2\xi\xi')^2 }}
\right]\xi' \, d\xi' \nonumber \\
&+& e^{2i\theta}\left( 
\frac{\partial^2}{\partial r^2}
-\frac{1}{r}\frac{\partial}{\partial r} 
\right)
\left[
\frac{2S^2_I(\tau+\sinh^{-1}u,\xi')-2S^2_{II}(\tau-\sinh^{-1}u,\xi')}{\sqrt{(\xi^2-\xi'^2-r^2)^2+(2\xi\xi')^2 }}-
\right]\xi' \, d\xi'
\nonumber \\
&+&\quad \textrm{etc.}
\label{eq:nonlocalized multipole angular distributions}
\end{eqnarray}
where
\[
u=\frac{\xi^2-\xi'^2-r^2}{2\xi\xi'}~.
\]
It is evident that each multipole term has its own distinguishing angular ($\theta$) and radial ($r$) dependence.

\section{RADIATION: PHYSICAL RELATION TO ITS SOURCE}

The radiation expressed by Eqs.(\ref{eq:multipole
expansion})-(\ref{eq:localized multipole angular distributions})
establishes the link between the accelerated sources and what is
measured and recorded in an expanding inertial reference frame.  The
measured observations yield very detailed knowledge about each
accelerated source (in $I$ and $II$) individually as well as about
their relation to each other.

\subsection{Fidelity}
\label{sec-fidelity:mathematical}

The most striking aspect of the radiation process is the fidelity of
the signal measured in $F$. To bring this fidelity into sharper
focus, consider the radiation from two localized multipole sources, one
localized at $\xi'=\xi'_{I}$ and the other at $\xi'=\xi'_{II}$. Their
$m$th multipole moments are therefore
\[
S^m_{I,II}(\tau',\,\xi')=S^m_{I,II}(\tau')
                                        \frac{\delta(\xi'-\xi'_{I,II})}{\xi'}\
~,
\]
Consequently, the multipole superposition has the form
\begin{eqnarray}
\psi_F(\xi,\tau,r,\theta)&=&\left[
\frac{2S^0_I(\tau+\sinh^{-1}u_I)}{\sqrt{(\xi^2-\xi_I'^2-r^2)^2+(2\xi\xi_I')^2 }}-
\frac{2S^0_{II}(\tau-\sinh^{-1}u_{II})}{\sqrt{(\xi^2-\xi_{II}'^2-r^2)^2+(2\xi\xi_{II}')^2 }}
\right] \nonumber \\
&-& e^{i\theta} \frac{\partial}{\partial r}
\left[
\frac{2S^1_I(\tau+\sinh^{-1}u_I)}{\sqrt{(\xi^2-\xi_I'^2-r^2)^2+(2\xi\xi_I')^2 }}-
\frac{2S^1_{II}(\tau-\sinh^{-1}u_{II})}{\sqrt{(\xi^2-\xi_{II}'^2-r^2)^2+(2\xi\xi_{II}')^2 }}
\right] \nonumber \\
&+& e^{2i\theta}\left( 
\frac{\partial^2}{\partial r^2}
-\frac{1}{r}\frac{\partial}{\partial r} 
\right)
\left[
\frac{2S^2_I(\tau+\sinh^{-1}u_I)}{\sqrt{(\xi^2-\xi_I'^2-r^2)^2+(2\xi\xi_I')^2 }}-
\frac{2S^2_{II}(\tau-\sinh^{-1}u_{II})}{\sqrt{(\xi^2-\xi_{II}'^2-r^2)^2+(2\xi\xi_{II}')^2 }}
\right]
\nonumber \\
&+&\quad \textrm{etc.}
\label{eq:localized multipole angular distributions}
\end{eqnarray}
where
\[
u_{I,II}=\frac{\xi^2-\xi_{I,II}'^2-r^2}{2\xi\xi'_{I,II}}~.
\]
Compare Eq.(\ref{eq:multipole field}) with
Eq.(\ref{eq:localized multipole angular distributions}). The temporal evolution of every localized accelerated multipole source
\[
S^m_{I,II}(\tau',\xi') \quad m=0,\pm 1,\pm 2, \cdots
\]
displays itself with 100\% fidelity as the correspondingly measurable
amplitude
\[
S^m_{I,II}(\tau \pm\sinh^{-1}u_{I,II}) \quad m=0,\pm 1,\pm 2, \cdots
\]
on the hypersurface of synchronous time $\xi=const$ of the expanding
inertial observation frame in $F$. There is no distortion and no
spatial chirp ($\tau$-dependent redshift), regardless how violently the localized
multipole source got accelerated.

As pointed out in section \ref{sec-fidelity:physical}, the high
fidelity is due to the \emph{expanding} nature of the \emph{inertial
observation frame}. Such a frame consists of an expanding set of free
float recording clocks with radio receivers all synchronized and
coherently phased to measure the complex amplitude (magnitude and
phase) of the spatial amplitude profile at any fixed synchronous time
$\xi >0$. Once these recording clocks have been brought into existence,
they can always be used to measure, receive, and record the e.m. field with 
100\% fidelity. 

Not so for the usual \emph{static inertial observation frame}, which
consist of a static lattice of free float meter rods,
clocks, and radio receivers.  Such a frame would be entirely
unsuitable for observing the emission of radiation from violently
accelerated bodies. Once the recording clocks have been assembled by
the physicist/observer into such a frame, the reception, measurement,
and recording of electromagnetically encoded information will always
be compromised by the destructive blueshift from the accelerated
source. 

\subsection{Spatial Structure of the Source}

The second striking feature of the emitted radiation is that its
measurement yields the spatial multipole structure of the
source. Measure the angular distribution for a given radial coordinate
$r$ in the plane transverse to the direction of acceleration. Do a
least squares fit to the measured data points in order to determine
each of the Fourier coefficients in Eq.(\ref{eq:multipole angular
distributions}). In order to obtain the radial distribution of each of
these coefficients, repeat this determination for various $r$ values.  A
second least squares analysis yields the radial derivatives and hence
the amplitude of each multipole moment $S^m_{I,II},~m=0,\pm 1,\pm 2,
\cdots$ in Eq.(\ref{eq:localized multipole angular distributions}).

\subsection{Double Slit Interference}
\label{sec-double slit}
The third property of the radiation process is that it highlights the
interference between the waves coming from Rindler sectors $I$ and
$II$. The interference pattern, which is recorded on a hypersurface of
synchronous time $\xi=constant$, has fringes whose separation 
yields the separation between the two localized in $I$ and $II$.
Let these sources be located symmetrically at 
\[
\xi'_I=\xi'_{II}\equiv \xi'_0~,
\]
and let them have equal proper frequency $\omega_0$ and hence (in
compliance with the first term of the wave Eq.(\ref{eq:full scalar
equation in I and II}) equal Rindler coordinate frequency
\[
\omega=\omega_0\xi'_0~.
\]
Consequently, they are characterized by their amplitudes and their
phases. Indeed, their form is 
\begin{eqnarray}
S^m_{I}(\tau +\sinh^{-1} u_I)&=&A^0_I \cos
	[\omega_0\xi'_0(\tau+\sinh^{-1}u_I)+\delta^m_I]\nonumber\\
S^m_{II}(\tau -\sinh^{-1} u_{II})&=&A^0_{II} \cos
	[\omega_0\xi'_0(\tau-\sinh^{-1}u_I)+\delta^m_{II}]~.
\label{eq:two sources}
\end{eqnarray}
Thus the full scalar field, Eq.(\ref{eq:localized multipole angular
distributions}), expresses two waves. Both propagate in the expanding
inertial frame, which is coordinatized by $(\xi,\tau,r,\theta)$. Their
respective wave crests are located in compliance with the constant
phase conditions $\tau\pm \sinh^{-1}u_I=const.$ Consequently, one wave
travels into the $+\tau$-direction with amplitude $A^m_{I}$, the other
into the $-\tau$-direction with amplitude $A^m_{II}$. They have well-determined phase velocities.
Together, these two waves form an interference pattern of standing waves,
\begin{eqnarray}
\psi_F(\xi,\tau,r,\theta)&=&
\frac{1}{\sqrt{(\xi^2-\xi_0'^2-r^2)^2+(2\xi\xi_0')^2 }}
\left[
	\begin{array}{c}
	~\\
	~
	\end{array}
(A^0_I-A^0_{II}) \cos [\omega_0\xi'_0(\tau-\sinh^{-1}u_I)+\delta^0_I]\right. \nonumber\\
&~&\quad\quad\quad\quad\quad\quad\quad\quad\quad\quad\quad\quad\quad-\left.
2A^0_{II} \sin \left(\omega_0\xi'_0\tau 
				+\frac{\delta^0_{II}+\delta^0_{I}}{2}\right) 
	\sin\left(\omega_0\xi'_0\sinh^{-1}u_I
				-\frac{\delta^0_{II}-\delta^0_{I}}{2}\right)
\right]\nonumber\\
&+&\quad \textrm{higher~multipole~terms~of~order~}m=1,2,3,\cdots \quad .
\label{eq:interfering amplitudes}
\end{eqnarray}
The amplitude of this interference pattern is $A^0_{II}>0$,
and there is a uniform background of amplitude $(A^0_I-A^0_{II})>0$.
At synchronous time $\xi$ the interference fringes along the $\tau$-direction
can be read off the factor 
\[
\sin (\omega_0\xi'_0\tau +\frac{\delta^0_{II}+\delta^0_{I}}{2})
\]
in Eq.(\ref{eq:interfering amplitudes}). Consequently, the fringes are
spaced by the amount
\[
\left(
\begin{array}{c}
	\textrm{proper}\\
	\textrm{fringe~spacing}	
\end{array}
\right)
=\frac{2\pi\xi}{\omega_0\xi'_0}=\frac{\xi}{\xi'_0}\times
\frac{1}{\left( \begin{array}{c}
			\textrm{proper~frequency}\\
			\textrm{of~the~source}
		\end{array} \right)}
=\frac{2\xi}{\textrm{source~separation}}\times 
\frac{1}{\left( \begin{array}{c}
			\textrm{proper~frequency}\\
			\textrm{of~ the~source}
			\end{array} \right)}~.
\]
This means that, analogous to a standard optical interference pattern,
the fringe spacing is inversely proportional to the distance $2\xi'_0$
between the two sources.  Furthermore, the position of this
interference pattern depends on the phase of source $I$ relative to
source $II$\cite{phase}. It is difficult to find a more welcome way
than the four Rindler sectors for double slit interference.

These observations lead to the conclusion that (i) the four Rindler
sectors quite naturally accommodate a \emph{double slit
interferometer}, and that (ii) the spatial as well as the temporal
properties of the interference fringes, together with the magnitude of
the travelling background wave, are enough to reconstruct every aspect
of the two sources, Eq.(\ref{eq:two sources}).

\subsection{The Rindler Interferometer}
\label{The Rindler Interferometer}

The double slit interferometer works as follows: A plane wave which
starts in Rindler sector $P$ gets split into two partial components
which propagate through $I$ and $II$. There they get modified by the
two respective scatterers.  They are two pointlike dipole loops
accelerating into opposite directions.  Each loop acts as a
transmitter which re-radiates the electromagnetic field from the
impinging wave. The e.m. fields emitted by these two transmitters exits through
the event horizons of $I$ and $II$. Upon recombining in $F$
they produce an interference pattern as measured on the hypersurface
$\xi=constant$ of the expanding inertial observation frame.  The
strength and the variations in this pattern are determined by (i) the
proper separation between the two scatterers, (ii) their relative
strengths and (iii) their relative phase. In fact, from this
interference pattern one can reconstruct the currents $\dot
q_I(\tau')$ and $\dot q_{II}(\tau')$, including the amplitudes, phases
for each of them.  In brief, the \emph{expanding} inertial observation
frame is the ``screen'' on which one can literally ``see'' what is
going on in each of the two accelerated frames $I$ and $II$.

\section{RADIATED POWER}
\label{sec-radiated power}

The electric and magnetic field components are obtained from the wave
function $\psi_F(\xi,\tau,r,\theta)$ by taking the partial derivatives
listed in the tables in Section II.C. With their help we shall now
find the Poynting vector component along the $\tau$-direction, namely
\[
\frac{1}{4\pi}(\hat B_r \hat E_\theta-\hat B_\theta \hat E_r)\xi
\equiv T^\xi_{~\tau}~.
\]
Its space integral,
\begin{equation}
\int_0^\infty \int_0^\infty \int _0^{2\pi} T^\xi_{~\tau}\,\xi d\tau
\, rdr \,d\theta ~,
\label{eq:tau momentum}
\end{equation}
is the total radiated momentum (=radiant energy flow) into the
$\tau$-direction. It is positive (resp. negative) whenever the source
is confined to Rindler sector $I$ (resp. $II$). Furthermore, the
$\tau$-momentum is independent of the synchronous time $\xi$ because
$\tau$ is a cyclic coordinate. This $\tau$-momentum measures the
energy radiated by the two accelerated sources, and it takes the place
of what in a static inertial reference frame is the emitted energy.

Both the T.E. and the T.M. field have the same Poynting vector
component along the $\tau$-direction. More precisely, reference to
the table of T.E. and the T.M. field components (Section II) shows that 
in Rindler sector $F$ this Poynting object is
\begin{equation}
T^\xi_{~\tau}=\frac{\xi}{4\pi}
\left[
	\frac{\partial}{\partial\xi}\left( \frac{\partial\psi}{\partial r}\right)
\frac{1}{\xi}\frac{\partial}{\partial\tau}
					\left( \frac{\partial\psi}{\partial r}\right)
+
\frac{\partial}{\partial\xi}
			\left( \frac{1}{r}\frac{\partial\psi}{\partial \theta}\right)
\frac{1}{\xi}\frac{\partial}{\partial\tau}
		\left( \frac{1}{r}\frac{\partial\psi}{\partial \theta}\right)
\right]
~,
\label{eq:stress tensor for T.E. and T.M. field}
\end{equation}
the same for both types of fields. Furthermore, the wave function
$\psi$ is governed by a wave equation, which is also common to both
fields.  Consequently, the mathematical analysis which relates
observations to the radiation sources is the same for both types of
radiation fields. However, it is the difference in two types of
sources which is important from the viewpoint of physics.

The only difference lies in the source and hence in the amplitude and
phase of $\psi$ in $F$. Comparing the ensuing Eq.(\ref{eq:magnetic
moment density}) with Eq.(\ref{eq:electric moment density}), one sees
that T.E. and T.M. polarized radiation are caused by the densities of
magnetic and electric dipole moment respectively.

\subsection{Axially Symmetric Source and Field}

The simplest nontrivial sources for the inhomogeneous wave
Eq.(\ref{eq:full scalar fields in I or II}) is those which are axially symmetric.
For T.E. radiation magnetic dipoles are the most important sources, while for T.M. radiation they are electric dipoles.

\subsubsection{Magnetic Dipole and its Radiation Field}

Consider radiation emitted from two magnetic dipoles. Have them be two
circular loop antennas each of area of $\pi a^2$ aligned parallel to
the $(x,y)$-plane with center on the $z$-axis. Fix their location in
Rindler sectors $I$ and $II$ by having them located at $\xi'=\xi'_I$
and $\xi'=\xi'_{II}$ so that they are accelerated into opposite
directions. Suppose each antenna has proper current
\[
i_{I,II}(\tau')=\frac{1}{\xi'_{I,II}} \frac{d~q_{I,II}(\tau')}{d\tau'} 
\quad\quad\quad\quad \left[\frac{\textrm{charge}}{\textrm{length}}\right]  ~.
\]
Then its magnetic moment is 
\[
\quad\quad\quad\quad\quad\quad\quad\quad i_{I,II}(\tau')\pi a^2
\equiv\textbf{m}_{I,II}(\tau')~,
\quad\quad\quad\quad \left[~\textrm{(proper
~current)}\times \textrm{(length)}^2=\textrm{dipole moment}~\right]
\]
its charge-flux four-vector obtained from Eq.(\ref{eq:polar T.E. source}) is
\begin{eqnarray}
\left( \hat S_{\tau'} ,\hat S_{\xi'} ,\hat S_{r'} ,\hat S_{\theta'}  \right)
 &=&\left(0,0,\frac{1}{r'} \frac{\partial S}{\partial \theta'},
			\frac{\partial S}{\partial r'} \right) \nonumber \\
			&=&i_{I,II}(\tau')\delta(\xi'-\xi'_{I,II})
 \delta(r' -a) 
	\left(0,0,0,-1\right)~,\quad\quad\quad\quad
				\left[\frac{\textrm{charge}}{\textrm{length}^3} \right]
\end{eqnarray}
and the corresponding scalar source function $S$ for Eq.(\ref{eq:full
scalar equation in I and II}) is
\begin{equation}
S:\quad S_{I,II}(\tau',\xi',r',\theta')= 
		\frac{d~q_{I,II}(\tau')}{d\tau'}\frac{\delta(\xi'-{\xi'}_{I,II})}{\xi'}\,
	\Theta (r'-a)
	\quad\quad\quad\quad \left[\frac{\textrm{charge}}{\textrm{length}^2}
								\right]~.
\label{eq:magnetic moment density}
\end{equation}
Here $\Theta$ is the Heaviside unit step function. The proper magnetic
dipole is the proper volume integral of this source,
\[
\int_0^\infty \int_0^\infty \int _0^{2\pi}
S_{I,II}(\tau',\xi',r',\theta') d\xi' r'dr' d\theta'=\pi a^2 i_{I,II}(\tau') 
						\quad\quad\quad\quad\quad
\left[~\textrm{(area)} \times \textrm{(proper current)}~\right]
\]
Being symmetric around its axis, such a source produces only radiation
which is independent of the polar angle $\theta$. Consequently, all
partial derivatives w.r.t. $\theta$ vanish, and the full scalar field,
Eq.(\ref{eq:localized multipole angular distributions}), in $F$
becomes with the help of Eq.(\ref{eq:multipole})
\begin{eqnarray}
\psi_F(\xi,\tau,r,\theta)=2\pi a^2 \!\!\!\!\!\!\!\!&&\left[ 
\frac{1}{\sqrt{(\xi^2-\xi^{'2}_I -r^2)^2+(2\xi \xi'_I)^2}}
\frac{dq_I(\tau+\sinh^{-1}u_I)}{d\tau} \right.\nonumber\\
&-&\left. \frac{1}{\sqrt{(\xi^2-\xi^{'2}_{II} -r^2)^2+(2\xi \xi'_{II})^2}}
\frac{dq_{II}(\tau-\sinh^{-1}u_{II})}{d\tau}
\right]~,\quad\quad [\textrm{charge}] 
\label{eq:field due to two localized T.E. sources}
\end{eqnarray}
where
\[
u_{I,II} =\frac{\xi^2-\xi^{'2}_{I,II} -r^2}{2\xi \xi'_{I,II}}
\]
This is the T.E. scalar field due to a localized pair of axially symmetric
loop antennas, each one with its own time dependent current. By
setting one of them to zero one obtains the radiation field due to the
other.
\subsubsection{Electric Dipole and its Radiation Field}

The most important electric dipole radiators are two linear antennas
each of length $a$ aligned parallel to the $z$-axis, located at
$\xi'=\xi'_I$ and $\xi'=\xi'_{II}$ located in Rindler sectors $I$ and $II$, 
and hence accelerated into opposite directions. Suppose each
antenna has electric dipole moment
\[
q_{I,II}(\tau')\,a\equiv\textbf{d}_{I,II}(\tau')\quad\quad\quad\quad\quad
\left[~\textrm{(charge)}\times \textrm{(length)}=\textrm{dipole moment}~\right]
\]
Its charge-flux four-vector obtained from Eq.(\ref{eq:polar
T.M. source}) is 
\begin{eqnarray}
\left( \hat S_{\tau'} ,\hat S_{\xi'} ,\hat S_{r'} ,\hat S_{\theta'}  \right)
 &=&
\left(
\xi\frac{\partial S}{\partial \xi'}, \frac{1}{\xi'}\frac{\partial S}{\partial \tau'},
0,0, \right) \nonumber \\
		&=&\left( q_{I,II}(\tau')a\frac{d}{d\xi'}\delta(\xi'-\xi'_{I,II}),
 \frac{1}{\xi'}\frac{d q(\tau')}{d\tau'}a \,\delta(\xi'-\xi'_{I,II}),0,0\right)
\frac{\delta(r'- 0)\delta(\theta'-\theta'_0)}{r'}
~,\quad	\left[\frac{\textrm{charge}}{\textrm{length}^3} \right]
\end{eqnarray}
and the corresponding scalar source function $S$ for Eq.(\ref{eq:full
scalar equation in I and II}) is
\begin{equation}
S:\quad S_{I,II}(\tau',\xi',r',\theta')=
	q_{I,II}(\tau')a~\delta(\xi'-\xi_{I,II})
				\frac{\delta(r'-0)\delta(\theta'-\theta'_0)}{r'}\,
\quad\quad\quad\quad
	\left[\frac{\textrm{charge}}{\textrm{length}^2} \right]~.
\label{eq:electric moment density}
\end{equation}
This source is symmetric around the $z$-axis because it is non-zero only at $r'=0$.
The electric dipole moment is the proper volume integral of this source,
\[
\int_0^\infty \int_0^\infty \int _0^{2\pi}
S_{I,II}(\tau',\xi',r',\theta') d\xi' r'dr' d\theta'=
q_{I,II}(\tau') a\quad\quad\quad\quad\quad
\left[~\textrm{(charge)} \times \textrm{ (length)}~\right]
\]
The axial symmetry of the source implies that its radiation is
independent of the polar angle $\theta$. Consequently, except some for a
source-dependent factor, the scalar field $\psi_F$ in $F$ is the same
as Eq.(\ref{eq:field due to two localized T.E. sources}).
One finds
\begin{eqnarray}
\psi_F(\xi,\tau,r,\theta)=2a \!\!\!\!\!\!\!\!&&\left[ 
\frac{\xi'_i}{\sqrt{(\xi^2-\xi^{'2}_I -r^2)^2+(2\xi \xi'_I)^2}}
~q_I(\tau+\sinh^{-1}u_I) \right.\nonumber\\
&-&\left. \frac{\xi'_i}{\sqrt{(\xi^2-\xi^{'2}_{II} -r^2)^2+(2\xi \xi'_{II})^2}}
~q_{II}(\tau-\sinh^{-1}u_{II})
\right]~.\quad\quad [\textrm{charge}] 
\label{eq:field due to two localized T.M. sources}
\end{eqnarray}
This is the T.M. scalar field due to a pair of localized linear
antennas, both situated on the $z$-axis, each one with its own
time-dependent dipole moment $q(\tau')\,a$. By setting one of them to
zero one obtains the radiation field due to the other.

\subsection{Flow of Radiant T.E. Field Energy}

Any loop antenna radiates only for a finite amount of
time. Consequently, one can calculate the flow of total emitted
energy, which is given by the spatial integral, Eq.(\ref{eq:tau
momentum}).  The fact that $\theta$ is a cyclic coordinate for axially
symmetric sources and radiation implies that $\hat E_r=\hat B_\theta
=0$ for the T.E. field. Consequently, the spatial integral, a
conserved quantity independent of time $\xi$, reduces with the help of
the table of derivatives in Section \ref{sec-Maxwell fields} to
\begin{eqnarray}
\int_{-\infty}^\infty \int_0^\infty \int _0^{2\pi} T^\xi_{~\tau}\,\xi d\tau
\, rdr \,d\theta 
&=&
\int_{-\infty}^\infty \int_0^\infty \int _0^{2\pi} 
\frac{1}{4\pi}\hat B_r \times\hat E_\theta \xi \,\xi d\tau \, rdr \,d\theta 
						\label{eq:first Pointing integral}\\
&=&
\frac{1}{4\pi}\int_0^\infty \int_0^\infty \int _0^{2\pi}
\frac{1}{\xi}\frac{\partial}{\partial r} \frac{\partial \psi_F}{\partial \tau}
\times \frac{\partial}{\partial r} \frac{\partial \psi_F}{\partial \xi}\xi
\,\xi d\tau \, rdr \,d\theta\nonumber\\
&=&(\pm)\frac{(\pi a^2)^2}{\xi'^4_{I,II}}\int_{-\infty}^\infty 
\frac{2}{3}
\left\{\left( \frac{d^3q_{I,II}(\tau)}{d\tau^3}\right) ^2
+\left( \frac{d^2q_{I,II}(\tau)}{d\tau^2}\right)^2
\right\}d\tau
\label{eq:Pointing integral}
\end{eqnarray}
The computation leading to the last line has been consigned to the
Appendix. This computed quantity is the total energy flow (energy
$\times$ velocity), or equivalently, the total momentum into the
$\tau$-direction, radiated by a magnetic dipole accelerated uniformly
in Rindler sector $I$ (upper sign) or in Rindler sector $II$ (lower
sign).

The full scalar radiation field
\begin{eqnarray}
\psi_F(\xi,\tau,r,\theta)=2\pi a^2 
\frac{\pm 1}{\sqrt{(\xi^2-\xi^{'2}_I -r^2)^2+(2\xi \xi'_I)^2}}
\frac{dq_{I,II}(\tau')}{d\tau}~,
\end{eqnarray}
with
\[
\tau'=\tau \pm \sinh^{-1}\frac{\xi^2-\xi^{'2}_{I,II} -r^2}{2\xi \xi'_{I,II}}~,
\]
is a linear functional correspondence. It maps the temporal history
$q_{I,II}(\tau')$ at $\xi'=\xi'_{I,II}$ in Rindler sector $I$
(resp. $II$) with 100\% fidelity onto a readily measurable e.m. field
along the $\tau$-axis (or a line parallel to it) on the spatial
hypersurface $\xi=const.$ in Rindler sector $F$. This correspondence
has 100\% fidelity because, aside from a $\tau$-independent factor,
the source history $q_{I,II}(\tau')$ and the scalar field
$\psi_F(\tau)$ differ only by a constant $\tau$-independent shift
on their respective domains $-\infty<\tau'<\infty$ and
$-\infty<\tau<\infty$. This implies that
\begin{equation}
\frac{\partial}{\partial\tau'}=\frac{\partial}{\partial\tau}
\label{eq:equal derivatives}
\end{equation}
when applied to the source function $q_{I,II}$. The expression for the
radiated momentum becomes more transparent physically if one uses
the proper time derivative
\[
\frac{d}{dt'} \equiv
\frac{1}{\xi'}\frac{\partial}{\partial\tau'}
					=\frac{1}{\xi'}\frac{\partial}{\partial\tau}
\]
at the source. Introduce the (proper) magnetic moment 
of the current loop having radius $a$:
\[
\textbf{m}=\pi a^2 \frac{1}{\xi'}\frac{\partial q}{\partial \tau'} ~.
\]
One finds from Eq.(\ref{eq:Pointing integral}) that the proper radiated
longitudinal momentum (i.e. physical, a.k.a. orthonormal, component of
energy flow pointing into the $\tau$-direction) measured per proper
spatial $\tau$-interval $\xi\, d\tau$ in $F$ is
\begin{equation}
{\mathcal I}_{T.E.}=
(\pm)~\frac{\xi'^2}{\xi^2} \frac{2}{3}\left[ 
			\left(\frac{d^2 \textbf{m}}{dt'^2}\right)^2
+\frac{1}{\xi'^2}\left(\frac{d \textbf{m}}{dt'}\right)^2 \right]~.
\label{eq:Larmor's formula}
\end{equation}
This is the formula for the proper radiant energy flow due to a magnetic dipole
moment subject to uniform linear acceleration $1/\xi'$. 
There are two factors of $1/\xi$. The the first converts the coordinate
$\tau$-momentum component into its physical component. The second is due
to the fact that Eq.(\ref{eq:Larmor's formula}) expresses this quantity 
per proper distance into the $\tau$ direction.

When the
acceleration $1/\xi'$ is small then the second term becomes small compared to
the first. In fact, one recovers the familiar Larmor
formula~\cite{Landau1962} relative to a \emph{static inertial} frame
by letting $\xi=\xi'$ and letting $1/\xi'\rightarrow 0$, which
corresponds to inertial motion. By contrast, Eq.(\ref{eq:Larmor's
formula}) is the correct formula for an \emph{accelerated} dipole
moment. However, observation of radiation from such a source entails
that the measurements be made relative to an \emph{expanding inertial}
reference frame.

\subsection{Flow of Radiant T.M. Field Energy}

The mathematical computation leading from Eq.(\ref{eq:first Pointing
integral}) and ending with Eq.(\ref{eq:Larmor's formula}) can be
extended without any effort to T.M. radiation. The extension consists of
replacing a T.E. source with a corresponding T.M. source,
\[
\frac{1}{\xi'}\frac{\partial q}{\partial\tau'}\pi a^2 \rightarrow qa~,
\]
or equivalently
\[
\textbf{m} \rightarrow \textbf{d}~.
\]
Consequently, the formula for the flow of T.M. radiant energy due to an
electric dipole subject to uniform linear acceleration $1/\xi'$ is 
\begin{equation}
{\mathcal I}_{T.M.}=
(\pm)~\frac{\xi'^2}{\xi^2} \frac{2}{3} \left[
                        \left(\frac{d^2 \textbf{d}}{dt'^2}\right)^2
+\frac{1}{\xi'^2}\left(\frac{d \textbf{d}}{dt'}\right)^2 \right]~.
\label{eq:Larmor's electric dipole formula}
\end{equation}
The justification for this extension is Eqs.(\ref{eq:stress tensor for
T.E. and T.M. field}) and (\ref{eq:equation for the unit impulse
response}).  They are the same for T.E. and T.M. radiation.  The
radiation intensity expressed by Eq.(\ref{eq:Larmor's electric dipole
formula}) extends the familiar Larmor formula for radiation from an
\emph{inertially moving} electric dipole~\cite{Landau1962} to one which is
\emph{accelerated} linearly and uniformly.

\section{VIOLENT ACCELERATION}
The second term in the radiation formula, Eq.(\ref{eq:Larmor's
formula}), is new. Under what circumstance does it dominate? Consider
the circumstance where the magnetic dipole oscillates with proper
frequency $\omega_0=2\pi/\lambda_0$. By averaging the emitted
radiation over one cycle, one finds
\[
(\pm)~\frac{\xi'^2}{\xi^2} \frac{2}{3}\left[
			\left(\frac{d^2 \textbf{m}}{dt'^2}\right)^2
+\frac{1}{\xi'^2}\left(\frac{d \textbf{m}}{dt'}\right)^2 \right]
\longrightarrow (\pm)~\frac{\xi'^2}{\xi^2} \frac{\omega_0^4}{3}\,\langle
\textbf{m}^2\rangle \left[1+  \frac{1}{\xi'^2 \omega_0^2} \right]
=(\pm)~\frac{\xi'^2}{\xi^2} \frac{\omega_0^4}{3}\,\langle
 \textbf{m}^2\rangle \left[1+  \frac{1}{(2\pi)^2}\frac{\lambda_0^2}{\xi'^2} \right]
\]
Thus the criterion for ``violent'' acceleration is that its inverse,
the Fermi-Walker length of the accelerated point object be small compared to the emitted wavelength,
\[
\frac{\lambda_0}{\xi'}\equiv \frac{\lambda_0\times (\textrm{proper~acc'n})}{c^2}
\gg 2\pi
\]
or equivalently
\begin{equation}
\frac{(\textrm{proper~acc'n})}{c}
\gg~\omega_0~.
\end{equation}
Recall that $c/$(proper acceleration) is the time it takes for the
oscillator to acquire a relativistic velocity relative to an inertial
frame. Also recall that $2\pi/\omega_0$ is the time for one oscillation cycle.
Consequently, the criterion for ``violence'' is that 
\begin{equation}
	\left(
	\begin{array}{c}
		\textrm{time for oscillator}\\
		\textrm{to acquire}\\
		\textrm{relativistic velocity}
	\end{array}
	\right) \ll
		(\textrm{oscillation period})~.
\label{eq:the criterion}
\end{equation}
When this condition is fulfilled, the Larmor contribution to the
radiation is eclipsed by the Rindler contribution.

\section{RADIATIVE VS. NONRADIATIVE MOMENERGY}

Consider a dipole moment, \textbf{m} or \textbf{d}, which is
time-independent in its own accelerated frame. The augmented Larmor
formula, Eq.(\ref{eq:Larmor's electric dipole formula}) and
(\ref{eq:Larmor's formula}), yields zero radiative
$\tau$-momentum relative the expanding inertial frame in Rindler
sector $F$:
\begin{equation}
\int_0^\infty \int_0^\infty \int _0^{2\pi} T^\xi_{~\tau}\,\xi d\tau
\, rdr \,d\theta =0~.
\label{eq:tau momentum equal to zero}
\end{equation}
However, for a non-zero static dipole moment the other
momenergy\footnote{The word \emph{momenergy}, a term first coined by
J.A. Wheeler, refers to the single concept which ordinarily is
referred to by the cumbersome word ``energy-momentum''. Given the fact
that, in the physics of particles and fields, the concepts ``momentum''
and ``energy'' are merely different aspects united by a change of
inertial frames into a new single concept, the case for
correspondingly uniting the compound word `` momentum-energy'', or
``energy-momentum'' into the single word ``momenergy'' is
appropriate.} components, also measured in $F$, are non-zero:
\begin{eqnarray}
\int_0^\infty \int_0^\infty \int _0^{2\pi} 
	T^\xi_{~\xi}\,\xi d\tau \, rdr \,d\theta 
		&\not =&0 \label{eq:xi momentum not zero}\\
\int_0^\infty \int_0^\infty \int _0^{2\pi}
        T^\xi_{~r}\,\xi d\tau \, rdr \,d\theta 
		&\not =&0 \label{eq:r momentum not zero}\\
\int_0^\infty \int_0^\infty \int _0^{2\pi}
        T^\xi_{~\theta}\,\xi d\tau \, rdr \,d\theta 
		& =&0~~~~~~~~~~~~~~~~~~~~~~~~~~(\textrm{``axial symmetry''}) 
\label{eq:theta momentum equal to zero}
\end{eqnarray} 
Equations (\ref{eq:tau momentum equal to zero})-(\ref{eq:theta
momentum equal to zero}) express an observationally and hence
conceptually precise distinction between the radiative and
non-radiative e.m. fields of a dipole source accelerated in Rindler
sector $I$: The augmented Larmor formula implies that the dipole emits
radiation if and only if its $\tau$-momentum, the spatial integral of
$T^\xi_{~\tau}$ in the expanding inertial frame, is non-zero.
Furthermore, the existence of a dipole field, static in Rindler sector
$I$, is expressed by the non-vanishing of the other momenergy
components, Eqs.(\ref{eq:xi momentum not zero})-(\ref{eq:r momentum
not zero}).  Like the $\tau$-momentum, these components are also
measurable in the expanding inertial frame. If the dipole is not
static then the the emitted radiation gets tracked by the
$\tau$-momentum. In that case the other momenergy components play an
auxiliary role. They only track the sum of static dipole field and
the radiative field, not the separate contributions.

\section{UNIFYING PERSPECTIVE}
The four Rindler sectors lend themselves to a unifying perspective.
The context which makes this possible is the emission and observation
of radiation from a body accelerated linearly and uniformly. The
requirement that signals be transmitted with 100\% fidelity implies
that the spacetime arena for this radiation process consist of two
adjacent Rindler coordinatized sectors, such as $I$ and $F$ or $II$
and $F$.

The unifying perspective applied to two adjacent Rindler sectors is
brought into a particularly sharp focus by the augmented Larmor
formula, Eqs.(\ref{eq:Larmor's formula}) or (\ref{eq:Larmor's electric
dipole formula}). This is because the physical basis of this formula
is a radiation process which starts in one of the two Rindler sectors
and ends in the other. Indeed, the radiative longitudinal momentum
(longitudinal flow of radiative energy) observed and measured in the
expanding inertial frame in $F$ is expressed directly in terms of the
behaviour of the dipole source in the accelerated frame in $I$ (or
$II$).

The unifying perspective applies to all four Rindler sectors if one
considers a scattering process which starts in $P$ and ends in $F$. In
such a process an e.m. wave starts in Rindler sector $P$, splits into
two partial waves which cross the past event horizons and enter the
respective Rindler sectors $I$ and $II$. The partial wave in $I$ excites
the internal degree of freedom of the dipole oscillator accelerated in
$I$.  There the oscillations constitute a source for the scattered
radiation which propagates into $F$. The other partial wave, which
propagates through Rindler sector $II$, also reaches $F$. There the
resultant interference pattern is observed and measured. It is evident
that this interference pattern is made possible by the properties of
the four Rindler sectors combined, the Rindler interferometer of
Section \ref{The Rindler Interferometer}.

Thus both the Rindler interferometer and the augmented Larmor formula
provide a unifying perspective. It joins adjacent Rindler coordinate
charts into a single spacetime arena for radiation and scattering
processes from accelerated bodies.

This perspective is at variance with a philosophy
which seeks a particle-antiparticle definition in
non-rectilinear coordinate systems in flat spacetime
\cite{Padmanabhan}.

Such a philosophy typically focuses on one of the Rindler charts to the
exclusion of all the others. The application of quantum field theory 
to such a chart leads to the paradox of spurious particle production in 
flat spacetime. As a result quantum theory remains meaningfully invariant
only under a subset of classically allowed coordinate transformations.

A proposed solution is to disallow -- in quantum theory -- a large
class of coordinate transformations, such as those leading to Rindler
charts $F$ or $I$ \cite{Padmanabhan}.

However, the fault does not lie with these Rindler charts. Instead, it
lies with the underlying philosophy which seeks a definition of the
particle-antiparticle concept in one of the coordinate charts while
ignoring reference to the others. Such a selective focus does not
comply with, and hence is forbidden
by the unifying perspective implied by the augmented Larmor formula
and by the Rindler double-slit interferometer.

\section{CONCLUSION}
The subject of this article is the physics of accelerated frames.
The theme is: ``How does one observe and measure the properties of 
violently accelerated bodies?'' Answering this question has led to three 
results.

First of all, suppose one considers the transmission of e.m. signals
from a translationally accelerated transmitter to a receiver in an
inertial frame.  One finds that the signals can be transmitted without
any time dependent Doppler distortion and with 100\% fidelity
\emph{provided} the signals are received in an inertial frame which is
\emph{expanding}. A static inertial frame would not do.

Second, Larmor's radiation formula for the radiation intensity from a
dipole source gets augmented if that dipole is subjected to linear
uniform acceleration. Under this circumstance the total radiation
is the sum of the magnetic and electric dipole radiation,
\begin{equation}
{\mathcal I}_{total}=
(\pm)~\frac{\xi'^2}{\xi^2} \left\{ 
			\frac{2}{3}\left[ 
                        \left(\frac{d^2 \textbf{d}}{dt'^2}\right)^2
+\frac{1}{\xi'^2}\left(\frac{d \textbf{d}}{dt'}\right)^2 \right]
+
\frac{2}{3}\left[ 
			\left(\frac{d^2 \textbf{m}}{dt'^2}\right)^2
+\frac{1}{\xi'^2}\left(\frac{d \textbf{m}}{dt'}\right)^2 \right]
			\right\}~.
\label{eq:Larmor's magnetic and electric dipole formula}
\end{equation}
The amount of that augmentation becomes dominant
when the acceleration is so large that its ``Fermi-Walker'' length
($\xi'=c^2$/acceleration) is smaller than the wavelength of the emitted
radiation, or equivalently, when Eq.(\ref{eq:the criterion}) is fulfilled.

Finally, taking note of the fact that for every source accelerated to
the right there is a twin source accelerated to the left, suppose
these two sources are irradiated coherently. Then these twins together
with their concomitant expanding inertial reference frame form an
interferometer. More precisely, the four Rindler coordinatized sectors
form an interferometer, the Rindler interferometer.  Its geometrical
arena consists of the Rindler sectors $P\rightarrow(I,II)\rightarrow
F$ as exhibited in Figure \ref{fig:Rindler spacetime}. They
accommodate what in Euclidean space would be the components of an
optical interferometer with (i) $P$ serving as the beam splitter, (ii)
the spacetimes of the two accelerated frames $I$ and $II$ serving as
the two arms, and (iii) the spacetime $F$ serving as the beam
re-combiner. The interference pattern is recorded by the expanding
inertial frame in $F$.

\section{ACKNOWLEDGEMENT}

The author would like to thank Nirmala Prakash and Yuri Obukhov for
helpful remarks.

\section{APPENDIX: POTENTIAL, FIELD AND RADIATED MOMENTUM OF AN 
ACCELERATED POINT-LIKE MAGNETIC DIPOLE MOMENT}

Even though the focus of this article is on the T.E. radiation from an
axially symmetric circular dipole antenna, the mathematical structure
of T.M.  radiation is virtually the same for both. One merely has to
interchange the magnetic with the electric field components to obtain
one from the other. Furthermore, both are derived from a single scalar
which obeys the same inhomogeneous scalar wave Eq.(\ref{eq:the full
equation}). For T.E. radiation from a magnetic dipole localized at
$r'=0$ and $\xi'=constant$ in Rindler sector $I$ (upper sign) or $II$
(lower sign), this scalar is the simplified version of
Eq.(\ref{eq:field due to two localized T.E. sources}):
\begin{eqnarray}
\psi_F(\xi,\tau,r,\theta)=\frac{2\pi a^2}{2\xi\xi'} 
\frac{\pm 1}{\sqrt{u^2+1}}
\frac{dq(\tau \pm \sinh^{-1}u)}{d\tau} 
\label{eq:field due to a single localized source}
\end{eqnarray}
where
\[
u =\frac{\xi^2-\xi^{'2} -r^2}{2\xi \xi'}
\]
What are the T.E. field components and what is the Poynting integral,
Eq.(\ref{eq:first Pointing integral})? With some care the calculation
is reasonably straight forward. We exhibit the calculational 
path in this appendix because it probably is optimal and hence
also useful for the case of T.M. radiation.

\subsection{Vector Potential}

For T.E. radiation the components of the vector potential relative to 
the orthonormal basis are
\begin{eqnarray}
(\hat A_\xi,\hat A_\tau,\hat A_r,\hat A_\theta )&=& 
\left(
	0,0,\frac{-1}{r}
\frac{\partial\psi_F}{\partial \theta},\frac{\partial\psi_F}{\partial r}
	\right) \nonumber \\
&=&
\left(
	0,0,~~~~0~~~~,\frac{\pm4\pi a^2r}{(2\xi\xi')^2}
	\left( \frac{u}{(u^2+1)^{3/2}} \dot q \mp \frac{1}{u^2+1} \ddot q
	\right) \right)
\label{eq:vector potential}
\end{eqnarray}
where an over-dot refers to the partial derivative
\[
\dot q=\frac{\partial q (\tau\pm \sinh^{-1} u)}{\partial \tau} ~.
\]

\subsection{Field Components}

In compliance with the table of derivatives in Section \ref{sec-the T.E. field},
the relevant non-zero component of the electric field is
\begin{eqnarray}
\hat E_\theta &\equiv& \hat F_{\theta\xi}=\frac{1}{r}\left(
\frac{\partial A_\xi}{\partial \theta}-\frac{\partial
A_\theta}{\partial \xi} \right)\nonumber\\
&=&-\frac{\partial}{\partial \xi}\frac{\partial\psi_F}{\partial r}~,
\end{eqnarray}
which with the help of Eq.(\ref{eq:vector potential}) becomes 
\begin{eqnarray}
\hat E_\theta &=&-(\mp ) 4\pi a^2~~r\frac{\partial}{\partial \xi}
	\left[ \frac{1}{(2\xi\xi')^2} \left(
		\frac{u}{(u^2+1)^{3/2}}\dot q \mp \frac{1}{u^2+1} \stackrel{..}{q}
					\right)
	\right]\nonumber\\
&=& 2\pi a^2 (\alpha \dot q +\beta \ddot q +\gamma \stackrel{...}{q})~,
\label{eq:electric field}
\end{eqnarray}
where
\begin{eqnarray}
\alpha &=& \frac{\mp 2r}{(2\xi\xi')^2} 
	\left(
\frac{-3}{\xi}\frac{u}{(u^2+1)^{5/2}}+\frac{1}{\xi'}\frac{1-2u^2}{(u^2+1)^{5/2}}
	\right)\nonumber\\
\beta &=& \frac{-2r}{(2\xi\xi')^2}
        \left(
\frac{1}{\xi}\frac{2-u^2}{(u^2+1)^2}
					+\frac{3}{\xi'}\frac{u}{(u^2+1)^2}
        \right)\nonumber\\
\gamma &=& \frac{\mp 2r}{(2\xi\xi')^2}
        \left(
\frac{1}{\xi}\frac{u}{(u^2+1)^{3/2}}-\frac{1}{\xi'}\frac{1}{(u^2+1)^{3/2}}
        \right) ~.
\end{eqnarray}
Similarly the relevant non-zero magnetic field component is
\begin{eqnarray}
\hat B_r \equiv \hat F_{\theta\tau}&=&\frac{1}{r\xi}\left(
\frac{\partial A_\tau}{\partial \theta}-\frac{\partial A_\theta}{\partial \tau}
						\right)\nonumber\\
&=&-\frac{1}{r\xi}\frac{\partial}{\partial\tau}
		\left( r \frac{\partial\psi_F}{\partial r} \right)
=-\frac{1}{\xi}\frac{\partial}{\partial\tau} \frac{\partial\psi_F}{\partial r}~,
\end{eqnarray}
which with the help of Eq.(\ref{eq:vector potential}) becomes
\begin{equation}
\hat B_r= 2\pi a^2(\delta \ddot q +\epsilon \stackrel{...}{q}) ~,
\label{eq:magnetic field}
\end{equation}    
where
\begin{eqnarray}
\delta&=&
 \frac{\mp 2r}{(2\xi\xi')^2}\frac{1}{\xi}\frac{u}{(u^2+1)^{3/2}}\nonumber\\
\epsilon&=&
 \frac{2r}{(2\xi\xi')^2}\frac{1}{\xi}\frac{1}{u^2+1}~.
\end{eqnarray}

\subsection{Radiated Momentum}

Being generated by a circular loop, the density of radiated momentum
pointing into the $\pm \tau$ direction is independent of the polar
angle $\theta$. Consequently, that density's spatial integral,
Eq.(\ref{eq:Pointing integral}), reduces to  
\begin{eqnarray}
\int_{-\infty}^\infty \int_0^\infty \int _0^{2\pi} T^\xi_{~\tau}\,\xi d\tau
\, rdr \,d\theta
&=&
\frac{1}{4\pi}\int_{-\infty}^\infty \int_0^\infty \int _0^{2\pi}
\hat B_r \times\hat E_\theta \xi \,\xi d\tau \, rdr \,d\theta
\nonumber\\
&=&
\frac{(2\pi a^2)^2 2\pi}{4\pi}
\int_{-\infty}^\infty \int_0^\infty
(\delta \ddot q +\epsilon \stackrel{...}{q} )(\alpha \dot q +\beta \ddot q +\gamma
\stackrel{...}{q})
\xi^2d\tau\, r\, dr
\end{eqnarray}
The $\tau$-integration affects only the dotted factors, and they
vanish outside a sufficiently large $\tau$-interval, i.e. $\dot
q(\pm\infty)=\ddot q(\pm\infty)=0$. Consequently, integration by parts
yields
\[
\int_{-\infty}^\infty\ddot q \dot q
\,d\tau=\int_{-\infty}^\infty\stackrel{...}{q} \,\stackrel{..}{q}\,d\tau=0
\]
and
\[
\int_{-\infty}^\infty\stackrel{...}{q}\,\stackrel{.}{q}
\,d\tau=-\int_{-\infty}^\infty{\ddot q}^2 d\tau \ne 0
\]
Thus there remain only three non-zero terms in the integral,
\begin{equation}
\int_{-\infty}^\infty \int_0^\infty \int _0^{2\pi} T^\xi_{~\tau}\,\xi d\tau
\, rdr \,d\theta=
\frac{(2\pi a^2)^2 2\pi}{4\pi}\int_{-\infty}^\infty \int_0^\infty
\left(\epsilon\gamma {\stackrel{...}{q}}^2 +
(\beta\delta -\alpha\epsilon ) {\ddot q}^2 \, \right)
		\xi^2 d\tau\, r\, dr~.
\label{eq:integral with three terms}
\end{equation}
The coefficients of the squared terms are 
\begin{eqnarray}
\epsilon\gamma&=& \mp \frac{4r^2}{(2\xi\xi')^4} 
\left(
\frac{1}{\xi^2}
	\frac{u}{(u^2+1)^{5/2}}-
				\frac{1}{\xi\xi'}
						\frac{1}{(u^2+1)^{5/2}}
       \right)\label{eq:1st expression}\\
\beta\delta&=&
\mp\frac{4r^2}{(2\xi\xi')^4}
\left(
	-\frac{1}{\xi^2} \frac{(2-u^2)u}{(u^2+1)^{7/2}}
					-\frac{3}{\xi\xi'} \frac{u^2}{(u^2+1)^{7/2}}
\right)\label{eq:2nd expression}\\
\alpha\epsilon&=&\mp \frac{4r^2}{(2\xi\xi')^4}
\left(
      -\frac{3}{\xi^2} \frac{u}{(u^2+1)^{7/2}}+\frac{1}{\xi\xi'} \frac{1-2u^2}{(u^2+1)^{7/2}}
\right)\label{eq:3rd expression}
\end{eqnarray}
We now take advantage of the fact that the integral,
Eq.(\ref{eq:integral with three terms}), is independent of the
synchronous time $\xi$. This simplifies the evaluation of the integral
considerably because one may assume 
\[
1\ll\frac{\xi}{\xi'}
\]
without changing the value of the integral. The final outcome is that (i) in each of
the expressions, Eqs.(\ref{eq:1st expression})-(\ref{eq:3rd expression}),
only the last term contributes to the $r$-integral and (ii) the integral
assumes a simple mathematical form if one introduces 
\[
u=\frac{\xi^2-\xi^{'2} -r^2}{2\xi \xi'},\quad\quad du=-\frac{2r\,dr}{2\xi\xi'}
\]
as the new integration variable. With this scheme one has 
\begin{eqnarray}
\int_0^\infty \cdots \frac{r^2r\,dr}{(2\xi\xi')^2}&=&
\frac{1}{2} \int_{(\xi^2-\xi'^2)/2\xi\xi'}^{-\infty} \cdots \left(
			\frac{\xi}{2\xi'}-\frac{\xi'}{2\xi}-u
								\right)(-)du
\label{eq:limiting integral}
\end{eqnarray}
The to-be-used integrands have the form 
\[
\frac{1}{(u^2+1)^{n/2}},\quad \frac{u}{(u^2+1)^{n/2}}\quad\quad\quad\quad n=5,7,\cdots~,
\]
both of which are always less than one in absolute value, even when
they get multiplied by $u$. Consequently, one is perfectly justified
in saying that
\begin{eqnarray}
\int_0^\infty \cdots \frac{r^2r\,dr}{(2\xi\xi')^2}&\rightarrow&
\frac{\xi}{4\xi'}\int_{-\infty}^{\infty} \cdots du 
\quad\textrm{whenever} \quad \xi'\ll\xi ~.
\label{eq:final limiting integral}
\end{eqnarray}
Taking note that only the last terms of Eqs.(\ref{eq:1st
expression})-(\ref{eq:3rd expression}) give nonzero contribution, apply
the limiting form,
Eq.(\ref{eq:final limiting integral}), to evaluate the integral,
Eq.(\ref{eq:integral with three terms}). One finds that 
\begin{eqnarray}
\int_{-\infty}^\infty \int_0^\infty \int _0^{2\pi} T^\xi_{~\tau}\,\xi d\tau
\, rdr \,d\theta&=&
\frac{(2\pi a^2)^2 2\pi}{4\pi}
				\frac{\pm 4}{(2\xi')^2}
\frac{1}{4{\xi'}^2} \int_{-\infty}^\infty 
\left( \stackrel{...}{q}^2
	\int_{-\infty}^\infty \frac{du}{(u^2+1)^{5/2}}+
			\ddot q^2 \int_{-\infty}^\infty \frac{du}{(u^2+1)^{5/2}} 
\right)d\tau \nonumber\\
&=&
(\pi a^2)^2 \frac{\pm 1}{2\xi'^4}\int_{-\infty}^\infty 
\left( \stackrel{...}{q}^2
	\int_{-\pi/2}^{\pi/2}\cos^3 \phi~d\phi+
		\ddot q^2\int_{-\pi/2}^{\pi/2}\cos^3 \phi ~d\phi
\right)d\tau
\end{eqnarray}
The value of the integral
\[
\int_{-\pi/2}^{\pi/2}\cos^3 \phi~d\phi=\frac{4}{3}
\]
implies that the final result is
\begin{equation}
\int_{-\infty}^\infty \int_0^\infty \int _0^{2\pi} T^\xi_{~\tau}\,\xi d\tau
\, rdr \,d\theta=
(\pm)\frac{(\pi a^2)^2}{\xi'^4}\int_{-\infty}^\infty 
\left(
	\frac{2}{3}\stackrel{...}{q}^2 +\frac{2}{3}\ddot q^2
\right)d\tau~,
\end{equation}
the total momentum into the $\tau$-direction radiated by a
magnetic dipole accelerated in Rindler sector $I$ (upper sign) or in
Rindler sector $II$ (lower sign). This is the result stated by
Eq.(\ref{eq:Pointing integral})


\begin{references}
\bibitem{detectors}Instead of meter sticks with equally spaced
markings, it is better to have equally spaced detectors and probes. This
means that instead of measuring a particle in relation to the
meterstick markings, measure it by the ``click'' this particle
produces in the corresponding detector.  Furthermore, by having the
clocks keep track of the relative phase between adjacent detection
probes, one can measure the wavelength of a electromagnetic wave train
passing through the lattice work of equally spaced probes.
\bibitem{Einstein1907b}The other breakthrough was, of course, his
focus on the ``principle of the uniqueness of free fall'' (``Eotvos
property''), which he identified as a manifestation of his
``equivalence hypothesis'' according to which a body's acceleration in
a static gravitational field is merely a kinematical effect due to
accelerated reference frame relative to which the body is observed.
\bibitem{HOW TO FIND IT}Simply express $G$ as a double Fourier
integral. Using Cauchy's integral formula, evaluate the first Fourier
(frequency) integral first. Do this by having the integration path go
slightly above the two poles on the real frequency axis. The result is
the sum of two contributions. This yields the sum of two Fourier integral
expressions for the zero order Hankel function of the first and second
kind respectively. Their sum is the zero order Bessel function.
\bibitem{factor of two} We have included a factor of two on the right
hand side of Eq.(\ref{eq:multipole}). We do this in order to guarantee
that when $m=0$ and when one considers a circular loop of area $\pi
a^2$ localized at $\xi'=\xi_{I,II}$, as in Eq.(\ref{eq:magnetic moment
density}), then the $d\xi'$-integral of Eq.(\ref{eq:multipole}) yields
the standard value $\int_0^\infty S_{I,II}(\tau',\xi')d\xi'=\pi
a^2 \times (proper~current)$ for the proper magnetic dipole moment.
\bibitem{phase}Remember that it is $-\delta^m_{II}$ which expresses a phase
increase because in $II$ $\tau$ increases towards the past.
\end{references}
\end{document}